\documentclass[onecolumn,journal,noadjust]{IEEEtran}
\usepackage{cite}

\ifCLASSINFOpdf
\else
\fi
\usepackage{amssymb}
\usepackage{dsfont}
\usepackage{float}
\usepackage{graphicx}
\newtheorem{lemma}{Lemma}

\newtheorem{definition}{Definition}

\newtheorem{theorem}{Theorem}
\newtheorem{remark}{Remark}
\newtheorem{corollary}{Corollary}
\usepackage{xcolor}
\usepackage{mathrsfs}
\usepackage{mathtools}
\usepackage{caption}
\usepackage[ colorlinks = true,              linkcolor = blue, urlcolor  =
blue,              citecolor = red,              anchorcolor = green,
]{hyperref}

\usepackage [autostyle, english = american]{csquotes}
\MakeOuterQuote{"}

\usepackage{amsmath}

\begin{document}
%

\title{Channel Coding for Gaussian Channels with\\Mean and Variance Constraints}
%
%
%

\author{%
  \begin{tabular}[t]{@{}c@{\hspace{6em}}c@{}}
    Adeel Mahmood & Aaron B.~Wagner \\
    Nokia Bell Labs & School of Electrical and Computer Engineering,\\
    & Cornell University
  \end{tabular}
}

\maketitle

\begin{abstract}
We consider channel coding for Gaussian channels with the recently introduced mean and variance cost constraints. Through matching converse and achievability bounds, we characterize the optimal first- and second-order performance. The main technical contribution of this paper is an achievability scheme which uses random codewords drawn from a mixture of three uniform distributions on $(n-1)$-spheres of radii $R_1, R_2$ and $R_3$, where $R_i = O(\sqrt{n})$ and $|R_i - R_j| = O(1)$. To analyze such a mixture distribution, we prove a lemma giving a uniform $O(\log n)$ bound, which holds with high probability, on the log ratio of the output distributions $Q_i^{cc}$ and $Q_j^{cc}$, where $Q_i^{cc}$ is induced by a random channel input uniformly distributed on an $(n-1)$-sphere of radius $R_i$. To facilitate the application of the usual central limit theorem, we also give a uniform $O(\log n)$ bound, which holds with high probability, on the log ratio of the output distributions $Q_i^{cc}$ and $Q^*_i$, where $Q_i^*$ is induced by a random channel input with i.i.d. components.                      
\end{abstract}

\begin{IEEEkeywords}
Channel coding, Gaussian channels, second-order coding rate, random coding, mixture distribution
\end{IEEEkeywords}

%

\IEEEpeerreviewmaketitle

\section{Introduction}

The two common forms of cost (or power) constraints in channel coding have been the maximal cost constraint specified by $c(\mathbf{X}) \leq \Gamma$ almost surely, or the expected cost constraint specified by $\mathbb{E}\left [ c(\mathbf{X}) \right] \leq \Gamma$, where $\mathbf{X} \in \mathbb{R}^n$ is a random channel input vector and $c(\cdot)$ is an additively separable cost function defined as 
\begin{align*}
    c(\mathbf{X}) \coloneqq \frac{1}{n} \sum_{i=1}^n c(X_i). 
\end{align*}

Recent works \cite{10619384} and \cite{mahmood2024improvedchannelcodingperformance} introduced the mean and variance (m.v.) cost constraint, specified by 
\begin{align}
\begin{split}
    \mathbb{E}\left [c(\mathbf{X}) \right] &\leq  \Gamma\\
    \text{Var}\left ( c(\mathbf{X}) \right) &\leq \frac{V}{n},
\end{split}
     \label{43v} 
\end{align}
for discrete memoryless channels (DMCs). With a variance parameter $V$, the mean and variance (m.v.) cost constraint generalizes the two existing frameworks in the sense that $V \to 0$ recovers the first- and second-order coding performance of the maximal cost constraint \cite{10619384} and $V \to \infty$ recovers the expected cost constraint. Beyond generalization, the m.v. cost constraint for $0 < V < \infty$ is shown to have practical advantages over both prior cost models. Unlike the maximal cost constraint, it allows for an improved second-order coding performance with feedback (\cite[Theorem 3]{10619384}, \cite[Theorem 2]{mahmood2024improvedchannelcodingperformance}). In particular, for DMCs with a unique capacity-cost-achieving distribution, second-order coding performance improvement via feedback is possible if and only if $V > 0$. Even without feedback, the coding performance under the m.v. cost constraint is superior (\cite[Theorem 1]{mahmood2024improvedchannelcodingperformance}). Specifically, if $r(V)$ and $r(0)$ denote the optimal non-feedback second-order coding rates under the m.v. and maximal cost constraints, respectively, then [2, Theorem 1] showed that $r(V) > r(0)$ for all $V > 0$.

Unlike the expected cost constraint, the m.v. cost constraint enforces a controlled, "ergodic" use of transmission power \cite[(5)]{mahmood2024improvedchannelcodingperformance}. This is essential for several practical reasons such as operating circuitry in the linear regime, minimizing
power consumption, and reducing interference with other terminals. The m.v. cost constraint allows the power to fluctuate above the threshold in a manner consistent with a noise process, thus making it a more realistic and natural cost model in practice than the restrictive maximal cost constraint.

The aforementioned performance improvements under the m.v. cost constraint were shown for DMCs only, although the ergodic behavior enforced by the m.v. cost constraint is generally true. In this paper, we study additive white Gaussian noise (AWGN) channels with the mean and variance cost constraint with the cost function taken to be $c(x) = x^2$. Specifically, we characterize the optimal first- and second-order coding rates (SOCR) under the m.v. cost constraint for AWGN channels subject to a non-vanishing average probability of error $\epsilon > 0$. Simultaneously, we also characterize the optimal average error probability as a function of the second-order rate with the baseline first-order rate fixed as the capacity-cost function \cite[(9.17)]{Cover2006}. The latter is motivated by the fact, which itself follows from our results, that the strong converse \cite[p. 208]{Cover2006} holds under the m.v. cost constraint. Concretely, we show that the asymptotic expansion of the maximum achievable rate $R^*(n, \epsilon, \Gamma, V)$ under the m.v. cost constraint is given by
\begin{align}
    R^*(n, \epsilon, \Gamma, V) = C(\Gamma) + O \left( \frac{1}{\sqrt{n}} \right),  \label{3m}   
\end{align}
where $C(\Gamma)$ denotes the capacity-cost function. The fact that the first-order term $C(\Gamma)$ is independent of $\epsilon \in (0, 1)$ is an interesting finding since the strong converse does not hold for AWGN channels subject to expectation-only constraint \cite[Theorem 77]{Polyanskiy2010}. However, if one considers maximal probability of error instead of average probability of error, then the strong converse holds for both m.v. cost constraint and expectation-only constraint \cite[(18)]{7156144}. In this paper, we only focus on the average error probability and non-feedback framework.

Another interesting observation about $(\ref{3m})$ is that the first-order term $C(\Gamma)$ is independent of the variance parameter $V$ for all finite $V \geq 0$. Instead, the variance parameter along with the channel dispersion plays a role in the second-order term $O(1/\sqrt{n})$, whose exact characterization is discussed later in the paper. More results on AWGN channels can be found in \cite{7282467} and \cite{8012458}.

It is also to be noted that the asymptotic expansion in $(\ref{3m})$ is proven for a \emph{fixed} finite variance parameter $V \geq 0$ as $n \to \infty$. Therefore, while $V \to \infty$ in $(\ref{43v})$ recovers the expectation-only constraint, we have 
\begin{align}
    C(\Gamma) = \lim_{V \to \infty} \lim_{n \to \infty} R^*(n, \epsilon, \Gamma, V) \neq \lim_{n \to \infty} \lim_{V \to \infty} R^*(n, \epsilon, \Gamma, V) \stackrel{(a)}{=} C\left(\frac{\Gamma}{1 - \epsilon} \right), 
\end{align}
where the equality $(a)$ above can be found in \cite[Theorem 77]{Polyanskiy2010}. The achievability scheme in \cite[Theorem 77]{Polyanskiy2010} achieves 
$$\lim_{n \to \infty} \lim_{V \to \infty} R^*(n, \epsilon, \Gamma, V) \geq C\left(\frac{\Gamma}{1 - \epsilon} \right)$$
by assigning zero cost to an approximately $\epsilon$ fraction of the codewords while drawing the remaining codewords uniformly from a sphere of radius approximately equal to $\sqrt{\frac{n\Gamma}{1 - \epsilon}}$. Such a scheme has $\operatorname{Var}(c(\mathbf{X})) = O(1)$ and is disallowed under the m.v. cost constraint formulation in $(\ref{43v})$. Moreover, in contrast to the $O(1/\sqrt{n})$ second-order term in $(\ref{3m})$, we have 
\begin{align}
    R^*(n, \epsilon, \Gamma, \infty) = C\left(\frac{\Gamma}{1 - \epsilon} \right) + O\left(\sqrt{\frac{\log n}{n}} \right)
\end{align}
from \cite[Theorem 1]{7156144}.  

Hence, the analysis under the m.v. constraint with a finite variance parameter $V \geq 0$ requires different techniques. Our achievability proof relies on a random coding scheme where the channel input has a distribution supported on at most three concentric spheres of radii $R_1$, $R_2$ and $R_3$, where $R_i = O(\sqrt{n})$ and $|R_i - R_j| = O(1)$. Each uniform distribution on a sphere induces a distribution $Q_i^{cc}$ on the channel output, and a mixture of such input uniform distributions induces a mixture distribution of $Q_1^{cc}, Q_2^{cc}$ and $Q_3^{cc}$ on the output. For $\mathbf{Y} \sim Q_i^{cc}$, we have that $\mathbb{E}[||\mathbf{Y}||^2] = O(n)$ and $||\mathbf{Y}||^2$ concentrates in probability over a set encompassing $\pm \sqrt{n}$ deviations around its mean. Furthermore, the probability density function $Q_i^{cc}$ is given in terms of the modified Bessel function of the first kind. To assist in the analysis of this mixture distribution, we use a uniform asymptotic expansion of the Bessel function followed by traditional series expansions to bound the $\log$ ratio of $Q_i^{cc}$ and $Q_j^{cc}$. Remarkably, the zeroth- and first-order terms, which are $O(n)$ and $O(\sqrt{n})$ respectively, cancel out, giving us an $O(\log n)$ bound. This $O(\log n)$ bound holds uniformly over a set on which $||\mathbf{Y}||^2$ concentrates. Moreover, to facilitate the application of the central limit theorem (CLT), we give a similar $O(\log n)$ bound on the $\log$ ratio of $Q_i^{cc}$ and $Q^*$, where $Q^*$ is the output distribution induced by an i.i.d. channel input. The discussion of this paragraph is formalized in Lemmas \ref{QccN}, \ref{Qccratiolemma} and \ref{Logn_lemma} in Section \ref{main_achievability_theorem}. The achievability theorem and its proof are also given in Section \ref{main_achievability_theorem}. 

For the proof of the matching converse result, the main technical component involves obtaining convergence of the distribution of the normalized sum of information densities
to a standard Gaussian CDF. Let $\mathbf{Y} = (Y_1, \ldots, Y_n)$ denote the random channel output when the input to the AWGN channel, denoted by $W$, is some fixed vector $\mathbf{x} = (x_1, \ldots, x_n)$. The sum of the information densities is given by 
\begin{align}
    \sum_{i=1}^n \log \frac{W(Y_i|x_i)}{Q^*(Y_i)}. \label{sphxsymm} 
\end{align}
A well-known observation is that the distribution of $(\ref{sphxsymm})$ depends on $\mathbf{x}$ only through its average power $c(\mathbf{x}) = ||\mathbf{x}||^2/n$ (see Lemma \ref{sphericalsymm}). Unlike the maximal cost constraint, $c(\mathbf{x})$ is not uniformly bounded in the mean and variance cost constraint framework. Hence, we prove that the normalized sum converges uniformly in distribution to the standard Gaussian, where the convergence is only uniform over typical vectors $\mathbf{x} \in \mathbb{R}^n$. For atypical vectors, we use standard concentration arguments. Section \ref{main_converse_section} is devoted to the converse theorem and its proof.

\section{Preliminaries}

We write $\mathbf{x} = (x_1, \ldots, x_n)$ to denote a vector and $\mathbf{X} = (X_1, \ldots, X_n)$ to denote a random vector in $\mathbb{R}^n$. For any $\mu \in \mathbb{R}$ and $\sigma^2 > 0$, let $\mathcal{N}(\mu, \sigma^2)$ denote the Gaussian distribution with mean $\mu$ and variance $\sigma^2$. Let $\Phi$ denote the standard Gaussian CDF. Let $\chi^2_n(\lambda)$ denote the noncentral chi-squared distribution with $n$ degrees of freedom and noncentrality parameter $\lambda$. If two random variables $X$ and $Y$ have the same distribution, we write $X \stackrel{d}{=} Y$.  The modified Bessel function of the first
kind of order $\nu$ is denoted by $I_{\nu}(x)$. We will write $\log$ to denote logarithm to the base $e$ and $\exp(x)$ to denote $e$ to the power of $x$. Define $S^{n-1}_R$ to be the $(n-1)$-sphere of radius $R$ centered at the origin, i.e., $S^{n-1}_R \coloneqq \{ \mathbf{x} \in \mathbb{R}^n : ||\mathbf{x}|| = R \}$. For any set $A \subset \mathbb{R}^n$, $\mathds{1}_A(\cdot) : \mathbb{R}^n \to \{0, 1 \}$ will denote the indicator function. Let $\mathcal{P}(\mathbb{R}^n)$ denote the set of all probability distributions over $\mathbb{R}^n$. If $P \in \mathcal{P}(\mathbb{R}^n)$ is an $n$-fold product distribution induced by some $P' \in \mathcal{P}(\mathbb{R})$, then we write 
\begin{align}
    P(\mathbf{x}) = \prod_{i=1}^n P'(x_i) = P'(\mathbf{x}) \label{abuse}
\end{align}
with some abuse of notation.

The additive white Gaussian noise (AWGN) channel models the relationship between the channel input $\mathbf{X}$ and output $\mathbf{Y}$ over $n$ channel uses as $\mathbf{Y} = \mathbf{X} + \mathbf{Z}$, where $\mathbf{Z} \sim \mathcal{N}(\mathbf{0}, N \cdot I_n)$ represents independent and identically distributed (i.i.d.) Gaussian noise with variance $N > 0$. The noise vector $\mathbf{Z}$ is independent of the input $\mathbf{X}$. For a single time step, we use $W$ to denote the conditional probability distribution associated with the channel, e.g., 
$W(\cdot | x) \coloneqq \mathcal{N}(x, N)$ for the AWGN channel. Also, $W(\mathbf{y} | \mathbf{x}) = \prod_{i=1}^n W(y_i | x_i)$ similar to the notation in $(\ref{abuse})$. For $\mathbf{X} \sim P$, we use $P \circ W$ to denote the joint probability distribution and $PW$ to denote the induced output distribution, i.e., $(\mathbf{X}, \mathbf{Y}) \sim P \circ W$ and $\mathbf{Y} \sim PW$. To denote the probability of an event $A$ involving $(\mathbf{X}, \mathbf{Y}) \sim P \circ W$, we will write $(P \circ W)(A)$; for example, 
\begin{align*}
    (P \circ W)\left( \|\mathbf{X}\|^2 + \|\mathbf{Y}\|^2 \geq 1 \right) &\coloneqq \int d(P \circ W)(\mathbf{x}, \mathbf{y})  \mathds{1}_{\{ \|\mathbf{x}\|^2 + \|\mathbf{y}\|^2 \geq 1\}}\left( \mathbf{x}, \mathbf{y}  \right),\\
    P\left( \|\mathbf{X} \|^2 \leq 2 \right) &\coloneqq \int dP(\mathbf{x}) \mathds{1}_{\{\|\mathbf{x} \|^2 \leq 2 \}}(\mathbf{x}).  
\end{align*}

Let the cost function $c : \mathbb{R} \to [0, \infty )$ be given by $c(x) = x^2$. For a channel input sequence $\mathbf{x} \in \mathbb{R}^n$, 
\begin{align*}
    c(\mathbf{x}) =  \frac{1}{n} \sum_{i=1}^n c(x_i) = \frac{||\mathbf{x}||^2}{n}.
\end{align*}

Given the m.v. cost constraint specified by the mean parameter $\Gamma$ and the variance parameter $V$, the capacity-cost function $C(\Gamma)$ of only the mean parameter $\Gamma$ plays a crucial role in the optimal first- and second-order coding rate expressions that we derive. For $\Gamma > 0$, we have  
\begin{align}
    C(\Gamma) \coloneqq \max_{P: \mathbb{E}_P[c(X)] \leq \Gamma} I(P, W), \label{defcc}
\end{align}
where 
\begin{align*}
    \mathbb{E}_P\left [c(X) \right ] = \int_{\mathbb{R}} x^2 dP(x).  
\end{align*}

\begin{definition}
    We use $P^*$ to denote the capacity-cost-achieving distribution in $(\ref{defcc})$ and $Q^* = P^*W$ to denote the induced output distribution. We define 
    \begin{align*}
    \nu_{x} &\coloneqq \text{Var}\left( \log \frac{W(Y|x)}{Q^*(Y)} \right),\quad  \text{ where } Y \sim W(\cdot|x),\\
    V(\Gamma) &\coloneqq \int_{\mathbb{R}} \nu_x P^*(x) dx.
\end{align*}
    
\end{definition}

\begin{lemma}
We have $P^* = \mathcal{N}(0, \Gamma)$ and $Q^* = \mathcal{N}(0, \Gamma + N)$. Thus, the capacity-cost function is given by 
\begin{align*}
    C(\Gamma) &= \frac{1}{2} \log \left(1 + \frac{\Gamma}{N} \right).
\end{align*} 
Furthermore, for any $x \in \mathbb{R}$ and $Y \sim W(\cdot|x)$, 
    \begin{align}
   \mathbb{E}\left [ \log \frac{W(Y|x)}{Q^*(Y)} \right] &= C(\Gamma) - C'(\Gamma) \left(\Gamma - c(x) \right) \label{equsentimes} \\
    C'(\Gamma) &= \frac{1}{2(\Gamma + N)} \notag \\
    \nu_x &= \frac{\Gamma^2 + 2x^2N}{2\left(N + \Gamma \right)^2} \notag \\
    V(\Gamma) &= \frac{\Gamma^2 + 2\Gamma N }{2\left(N + \Gamma \right)^2}. \label{VGammadef}
\end{align}
\label{oftusedlemma}
\end{lemma}
\textit{Proof:} The fact that $P^* = \mathcal{N}(0, \Gamma)$ can be found in \cite[(9.17)]{Cover2006}. The remaining content of Lemma \ref{oftusedlemma} follows from elementary calculus.

\begin{lemma}[Spherical Symmetry]
    Let $\mathbf{x} \in \mathbb{R}^n$ be the channel input. Let $Y_i = x_i + Z_i$. Define 
    \begin{align*}
  T_i \coloneqq \log \frac{W(Y_i|x_i)}{Q^*(Y_i)} - \mathbb{E} \left [\log  \frac{W(Y_i|x_i)}{Q^*(Y_i)} \right ].
\end{align*}
Then  
$$\sum_{i=1}^n T_i \stackrel{d}{=} -\frac{\Gamma}{2(\Gamma + N)} \Lambda + \frac{N n c(\mathbf{x})}{2\Gamma(\Gamma + N)}  +  \frac{n\Gamma }{2(\Gamma + N)},$$
where $\Lambda \sim \chi^2_n\left(\frac{N n c(\mathbf{x})}{\Gamma^2} \right)$. Hence, the distribution of $\sum \limits_{i=1}^n T_i$ depends on $\mathbf{x}$ only through its cost $c(\mathbf{x})$. Hence, we can write 
\begin{align}
    \sum_{i=1}^n T_i \stackrel{d}{=} \sum_{i=1}^n \tilde{T}_i \label{redefTis}
\end{align}
where $\tilde{T}_i$'s are i.i.d. and 
\begin{align}
    \tilde{T}_i \stackrel{d}{=} \log \frac{W(Y|\sqrt{c(\mathbf{x})})}{Q^*(Y)} - \mathbb{E} \left [\log  \frac{W(Y|\sqrt{c(\mathbf{x})})}{Q^*(Y)} \right ], \label{v3}    
\end{align}
where $Y \sim \mathcal{N}(\sqrt{c(\mathbf{x})}, N)$ in $(\ref{v3})$.

\label{sphericalsymm}
\end{lemma}
\textit{Proof:} The observation of spherical symmetry with respect to the 
channel input is standard in most works on AWGN channels (see, e.g., \cite{5452208}). For convenience and completeness, we have given a proof of this lemma 
in Appendix \ref{sphericalsymmproof}. 

\begin{corollary}
    With the same setup as in Lemma \ref{sphericalsymm}, we have 
    \begin{align*}
        \text{Var}\left( \sum_{i=1}^n T_i \right) = \frac{n\Gamma^2 + 2Nn c(\mathbf{x})}{2(\Gamma + N)^2}. 
    \end{align*}
    \label{lem2cor}
\end{corollary}
\textit{Proof:} Use the equality in distribution given in $(\ref{redefTis})$ and Lemma \ref{oftusedlemma}.   

With a blocklength $n$ and a fixed rate $R > 0$, let $\mathcal{M}_R = \{1, \ldots, \lceil \exp(nR) \rceil \}$ denote the message set. Let $M \in \mathcal{M}_R$ denote the random message drawn uniformly from the message set.

\begin{definition}
An $(n, R)$ code for an AWGN channel consists of an encoder $f$ which, for each message $m \in \mathcal{M}_R$, chooses an input $\mathbf{X} = f(m) \in \mathbb{R}^n$, and a decoder $g$ which maps the output $\mathbf{Y}$ to $\hat{m} \in \mathcal{M}_R$. The code $(f,g)$ is random if $f$ or $g$ is random. 
\label{defwocostwofeedback}
\end{definition}

\begin{definition}
An $(n, R, \Gamma, V)$ code for an AWGN channel is an $(n, R)$ code such that $\mathbb{E}\left [ c(\mathbf{X}) \right] \leq \Gamma$ and $\text{Var}\left(c(\mathbf{X}) \right) \leq V/n$, where the message $M \sim \text{Unif}(\mathcal{M}_R)$ has a uniform distribution over the message set $\mathcal{M}_R$. 
\label{defwcostwofeedback2}
\end{definition}

Given $\epsilon \in (0, 1)$, define 
\begin{align*}
    M^*(n, \epsilon,  \Gamma, V) \coloneqq \max \{ \lceil \exp(nR) \rceil : \bar{P}_{\text{e}}(n,R, \Gamma, V) \leq \epsilon   \},
\end{align*}
where $\bar{P}_{\text{e}}(n,R,\Gamma, V)$ denotes the minimum average error probability over all random $(n,R, \Gamma, V)$ codes.

\begin{definition}
Define the function $\mathcal{K} : \mathbb{R} \times [0, \infty) \to (0, 1)$ as  
\begin{align}
    \mathcal{K}\left(r, V \right) &\coloneqq \inf_{\substack{P_\Pi \in \mathcal{P}(\mathbb{R}):\\
    \mathbb{E}[\Pi] = r \\
    \text{Var}(\Pi) \leq V
    }} \mathbb{E}\left [\Phi(\Pi) \right], \label{min5} 
\end{align}
where the infimum is over all random variables $\Pi$ (equivalently, probability distributions $P_{\Pi}$) that satisfy the given mean and variance constraints.
\label{Kfuncdef}
\end{definition}

\begin{remark}
    The $\mathcal{K}$ function originates from second-order coding rate analysis which entails a Gaussian approximation (using the central limit theorem) of the normalized information density. The mean and variance constraints in $(\ref{min5})$ are inherited from the block-level mean and variance constraints on the channel input vector $\mathbf{X}$.  
\end{remark}

\begin{lemma}
    The $\mathcal{K}$ function has the following properties: 
\begin{enumerate}
    \item The infimum in $(\ref{min5})$ is a minimum, and there exists a minimizer which is a discrete probability distribution with at most 3 point masses, 
    \item $\mathcal{K}(r, V)$ is (jointly) continuous in $(r, V)$, 
    \item for any fixed $V$, $\mathcal{K}(r, V)$ is strictly increasing in $r$,
    \item for any fixed $r$, $\mathcal{K}(r, V)$ is strictly decreasing in $V$,
    \item for any $V$ and $0 < \epsilon < 1$, there always exists a unique $r^*$ satisfying $\mathcal{K}(r^*, V) = \epsilon$, 
    \item for all $r \in \mathbb{R}$ and $V > 0$, we have $\mathcal{K}(r, V) < \Phi(r)$; therefore, the minimizing distribution in $(\ref{min5})$ has at least two point masses. 
\end{enumerate}
\label{Kfuncproperties}
\end{lemma}
\textit{Proof:} See \cite[Lemmas 3 and 4]{10619384} for properties $1,2, 3$ and $5$ and \cite[Theorem 1]{mahmood2024improvedchannelcodingperformance} for property $6$ above. For property 4, see Appendix \ref{prop4proof}.   

\begin{corollary}
    From Lemma \ref{Kfuncproperties}, an equivalent definition of $\mathcal{K}(r, V)$ for $V > 0$  is 
\begin{align}
    \mathcal{K}\left(r, V \right) &= \min_{\substack{P_\Pi \in \mathcal{P}(\mathbb{R}):\\
    \mathbb{E}[\Pi] = r \\
    \text{Var}(\Pi) = V\\
    2 \leq |\text{supp}(P_\Pi)| \leq 3
    }} \mathbb{E}\left [\Phi(\Pi) \right]. \label{vc2} 
\end{align}
\end{corollary}

\subsection{Numerical Simulation}

\begin{figure}[H]
    \centering
\includegraphics[width=10cm]{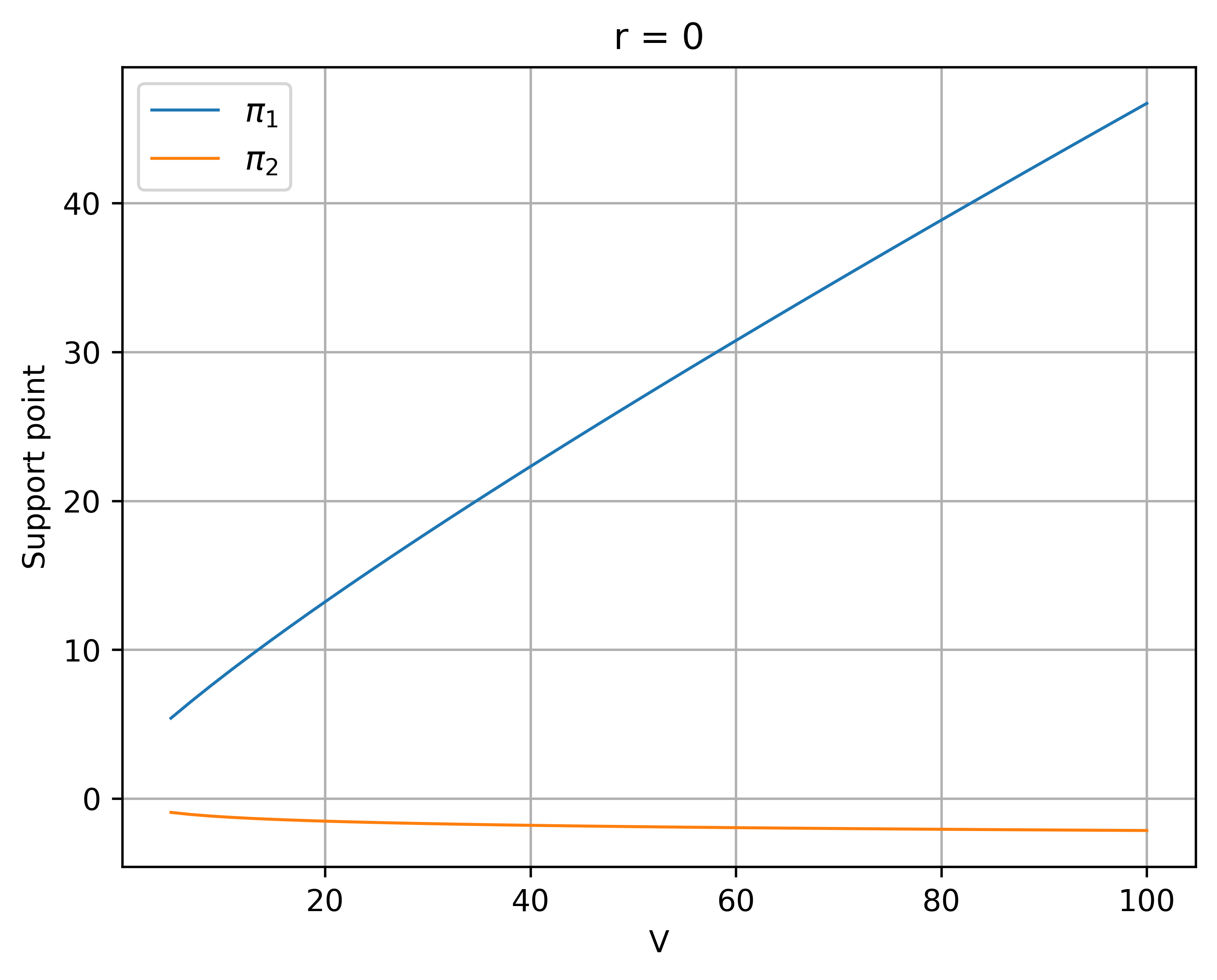}
\caption{}
\label{prob_locs}
\end{figure}

\begin{figure}[H]
    \centering
\includegraphics[width=10cm]{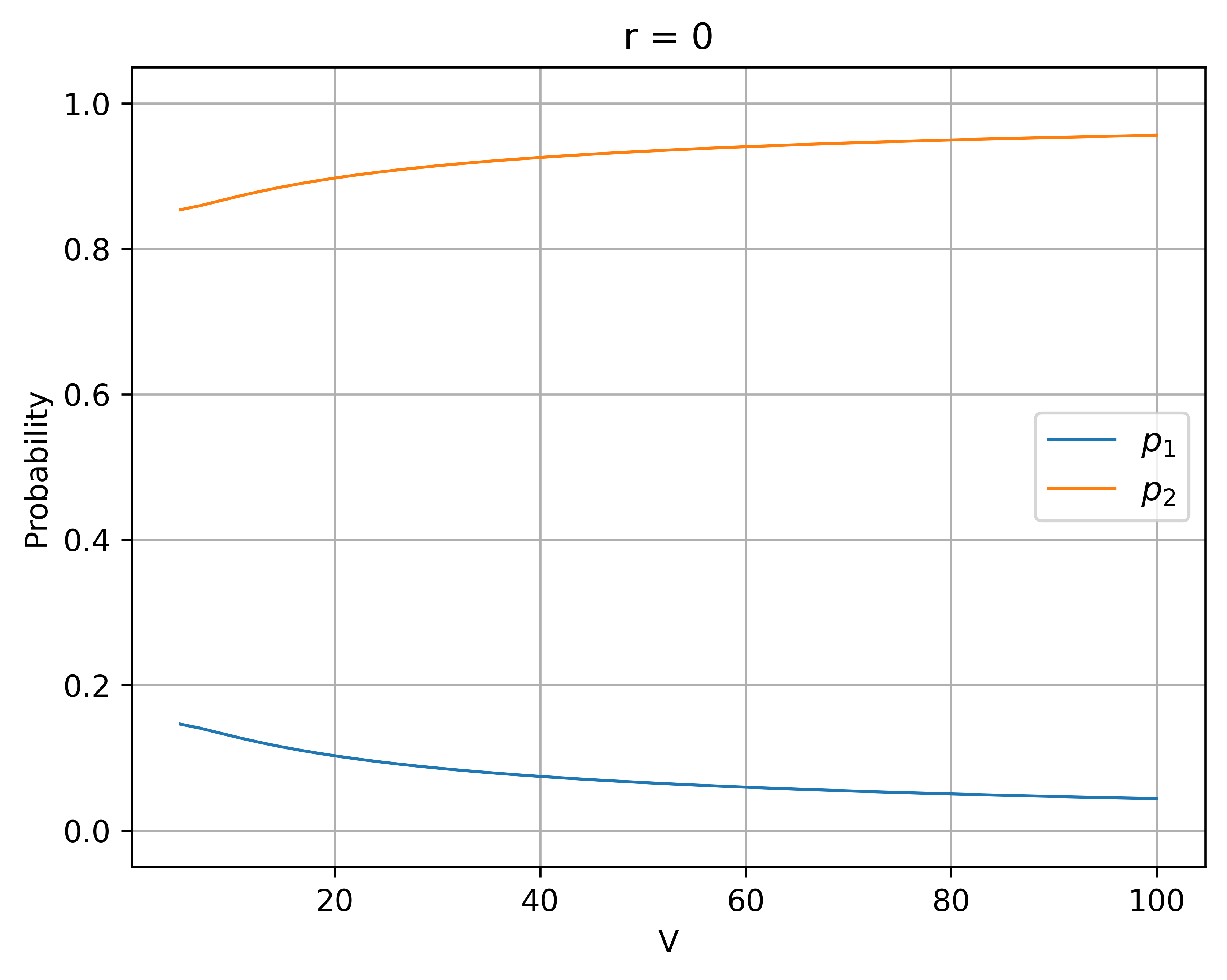}
\caption{}
\label{prob_assignments}
\end{figure}

Let 
\begin{align}
    P_{\Pi}(\pi) = \begin{cases}
    p_1 & \pi = \pi_1\\
    p_2 & \pi = \pi_2\\
    p_3 & \pi = \pi_3
    \end{cases}
    \label{minimizing_dist}
\end{align}
denote a minimizer in $(\ref{min5})$. In Fig. \ref{prob_locs} and Fig. \ref{prob_assignments}, we plot the support points $\pi_i$ and their respective probabilities $p_i$ versus the variance parameter $V$ for a fixed value of $r$. For the chosen range of values, the numerical solver produces a two-point distribution so only $(p_1, \pi_1)$ and $(p_2, \pi_2)$ are shown. This motivates the conjecture that in some regimes, a two-point mass probability distribution suffices to attain the minimum in $(\ref{min5})$ in which case a second-order optimal channel input distribution under the m.v. cost constraint would draw codewords from two power levels only, i.e., be a mixture distribution of two uniform distributions on $(n-1)$-spheres instead of three. 

As $V$ increases, the smaller support point $\pi_2$ decreases only slightly, while the larger support point $\pi_1$ increases substantially. Most of the probability is assigned to the smaller support point $\pi_2$ as shown in Fig. \ref{prob_assignments}. As will be seen later, such a two-point minimizer in Fig. \ref{prob_locs} and Fig. \ref{prob_assignments} will correspond to a coding scheme where the codewords are drawn with a low-power level $n \Gamma_1$ with low probability and a high-power level $n \Gamma_2$ with high probability, where the decreasing $\pi_i \mapsto  \Gamma_i$ map is given in $(\ref{38n})$ and the form of the mixture distribution is given in $(\ref{53vn})$.            

\section{Achievability Result \label{main_achievability_theorem}}

The design of random channel codes $(f,g)$ can be directly related to the design of the distribution $P \in \mathcal{P}(\mathbb{R}^n)$ of the codewords. Specifically, for each message $m \in \mathcal{M}_R$, the channel input $\mathbf{X}$ is chosen randomly according to $P$. Given an observed $\mathbf{y}$ at the decoder $g$, the decoder selects the message $m$ with the lowest index that achieves the maximum over $m$ of 
\begin{align*}
   \prod_{i=1}^n W\left (y_i | f(m) \right ). 
\end{align*}

Any distribution $P \in \mathcal{P}(\mathbb{R}^n)$ can be used to construct an $(n, R)$ channel code using the aforementioned construction. Moreover, any distribution $P$ satisfying
\begin{align}
\begin{split}
    \mathbb{E}\left [ c(\mathbf{X}) \right] &\leq \Gamma\\
    \text{Var}\left(c(\mathbf{X}) \right) &\leq \frac{V}{n} 
    \end{split}
    \label{faby}
\end{align}
can be used to construct a random $(n, R, \Gamma, V)$ code.

Given a random $(n ,R, \Gamma, V)$ code based on an input distribution $P$, the following lemma gives an upper bound on the average error probability of the code in terms of the distribution of $(\mathbf{X}, \mathbf{Y})$ induced by $P$ and the channel transition probability $W$.

\begin{lemma}
Consider an AWGN channel $W$ with noise variance $N > 0$ and cost constraint $(\Gamma,V) \in (0, \infty) \times [0, \infty)$. For any distribution $P \in \mathcal{P}(\mathbb{R}^n)$ satisfying $(\ref{faby})$, and any $n$, $\theta$ and $R$,
    \begin{align}
        \bar{P}_{\text{e}}(n,R,  \Gamma, V) \leq (P \circ W) \left(\frac{1}{n} \log \frac{W(\mathbf{Y}|\mathbf{X})}{PW(\mathbf{Y})} \leq R + \theta \right) + e^{-n \theta}, \label{patanahimujhe}
    \end{align}
    where $(\mathbf{X}, \mathbf{Y})$ have the joint distribution specified by 
    \begin{align*}
        (P \circ W)(\mathbf{x}, \mathbf{y}) &= P(\mathbf{x}) \prod_{k=1}^n W(y_k|x_k),
    \end{align*}
    and $PW$ denotes the marginal distribution of $\mathbf{Y}$. Furthermore, if for some $\alpha$ and $\epsilon$,
    \begin{align}
        \limsup_{n \to \infty} \, (P \circ W)\left(\frac{1}{n} \log \frac{W(\mathbf{Y}|\mathbf{X})}{PW(\mathbf{Y})} \leq C(\Gamma) + \frac{\alpha}{\sqrt{n}} \right) < \epsilon, \label{ghi}
    \end{align}
    then the distribution $P$ gives rise to an achievable second-order coding rate (SOCR) of $\alpha$, i.e.,  
    \begin{align}
        \liminf_{n \to \infty} \frac{\log M^*(n, \epsilon, \Gamma, V) - nC(\Gamma)}{\sqrt{n}} \geq \alpha. \label{ghi2}
    \end{align}
    \label{aaron'slemma}
    
\end{lemma}

\textit{Proof:} The proof can be adapted from the proof of \cite[Lemma 14]{9099482} by (i) replacing controllers with distributions $P$ satisfying $(\ref{faby})$ and (ii) replacing sums with integrals.

Lemma \ref{aaron'slemma} is a starting point for proving our achievability result. Hence, a central quantity of interest in the proof is
\begin{align}
  \log \frac{W(\mathbf{Y}|\mathbf{X})}{PW(\mathbf{Y})}. \label{centralinterest}  
\end{align}
As discussed in the introduction, Lemmas \ref{QccN}, \ref{Qccratiolemma} and \ref{Logn_lemma} are helpful in the analysis of $(\ref{centralinterest})$, where $P$ is a mixture distribution. Specifically, the lemmas are helpful in approximating the output distribution $PW$ in terms of a simpler product distribution.

\begin{lemma}
    Consider a random vector $\mathbf{Y} = \mathbf{X} + \mathbf{Z}$, where $\mathbf{X}$ and $\mathbf{Z}$ are independent, $\mathbf{X}$ is uniformly distributed on an $(n-1)$-sphere of radius $R$ and $\mathbf{Z} \sim \mathcal{N}(\mathbf{0}, N I_n)$. Let $Q^{cc}$ denote the PDF of $\mathbf{Y}$. Then 
    \begin{align*}
        Q^{cc}(\mathbf{y}) = \frac{\Gamma\left( \frac{n}{2} \right)}{2 (\pi N)^{n/2}}  \cdot  \exp\left( -\frac{R^2 + ||\mathbf{y}||^2}{2N} \right)  \left( \frac{N}{R||\mathbf{y}||}\right)^{\frac{n}{2}-1}  I_{\frac{n}{2}-1}\left(\frac{R||\mathbf{y}||}{N}\right).
    \end{align*}
    \label{QccN}
\end{lemma}

\textit{Proof:} The proof is given in Appendix \ref{QccNproof}.

To state the next two lemmas in a succinct way, we need to introduce some notation. 

\begin{definition}[Multi-parameter and multi-variable big-$O$ notation] Let \(f, f_1, \ldots, f_m\) be functions of $\epsilon, \Delta$ and $n$. We write  
\begin{align*}
    f(\epsilon, \Delta, n) = O\left( f_1(\epsilon, \Delta, n), \ldots, f_m(\epsilon, \Delta, n) \right) 
\end{align*}
if there exist positive constants $ \epsilon_0, \Delta_0, n_0$ and $C_1, \ldots, C_m$ such that for all $n \geq n_0$, $|\epsilon| \leq \epsilon_0$ and $|\Delta| \leq \Delta_0$,  
\begin{align*}
    |f(\epsilon, \Delta, n )| \leq  \sum_{i=1}^m C_i |f_i(\epsilon, \Delta, n)|.
\end{align*}
\end{definition}

\begin{lemma}
    Consider a random vector $\mathbf{Y} = \mathbf{X} + \mathbf{Z}$, where $\mathbf{X}$ and $\mathbf{Z}$ are independent, $\mathbf{X}$ is uniformly distributed on an $(n-1)$-sphere of radius $\sqrt{n \Gamma}$ and $\mathbf{Z} \sim \mathcal{N}(\mathbf{0}, N I_n)$. Let $Q^{cc}$ denote the PDF of $\mathbf{Y}$. Consider another random vector $\mathbf{Y}' = \mathbf{X}' + \mathbf{Z}'$, where $\mathbf{X}'$ and $\mathbf{Z}'$ are independent, $\mathbf{X}'$ is uniformly distributed on an $(n-1)$-sphere of radius $\sqrt{n \Gamma'}$ and $\mathbf{Z}' \sim \mathcal{N}(\mathbf{0}, N I_n)$. Let $Q^{cc}_0$ denote the PDF of $\mathbf{Y}'$. Let $\Gamma' = \Gamma + \epsilon$. Then  
    \begin{align*}
        \sup_{\mathbf{y} \in \mathcal{P}_n^*} \Bigg | \log \frac{Q^{cc}_0(\mathbf{y})}{Q^{cc}(\mathbf{y})} \Bigg | = O\left(\log n, n\epsilon^2, n\Delta^2, n\epsilon \Delta \right),
    \end{align*}
    where $\mathcal{P}_n^* = \left \{ \mathbf{y} \in \mathbb{R}^n: \Gamma + N - \Delta \leq  \frac{||\mathbf{y}||^2}{n} \leq \Gamma + N + \Delta  \right \}$ and $\Delta \geq 0$. 
\label{Qccratiolemma}
\end{lemma}

\begin{remark}
    The parameters $\epsilon$ and $\Delta$ in Lemma \ref{Qccratiolemma} may depend on $n$.
\end{remark}

\textit{Proof:} The proof of Lemma \ref{Qccratiolemma} is given in Appendix \ref{Qccratiolemmaproof}.

\begin{lemma}
    Consider a random vector $\mathbf{Y} = \mathbf{X} + \mathbf{Z}$, where $\mathbf{X}$ and $\mathbf{Z}$ are independent, $\mathbf{X}$ is uniformly distributed on an $(n-1)$ sphere of radius $\sqrt{n \Gamma}$ and $\mathbf{Z} \sim \mathcal{N}(\mathbf{0}, N I_n)$. Let $Q^{cc}$ denote the PDF of $\mathbf{Y}$ and let $Q^* = \mathcal{N}(\mathbf{0}, (\Gamma + N) I_n)$. Then  
    \begin{align*}
        \sup_{\mathbf{y} \in \mathcal{P}_n^*} \Bigg | \log \frac{Q^{cc}(\mathbf{y})}{Q^*(\mathbf{y})} \Bigg | = O \left( \log n, n \Delta^2 \right), 
    \end{align*}
    where $\mathcal{P}_n^* = \left \{ \mathbf{y} \in \mathbb{R}^n: \Gamma + N - \Delta  \leq  \frac{||\mathbf{y}||^2}{n} \leq \Gamma + N + \Delta  \right \}$ and $\Delta \geq 0$. 
    \label{Logn_lemma}
\end{lemma}
\begin{remark}
    The parameter $\Delta$ in Lemma \ref{Logn_lemma} may depend on $n$. In fact, in the proof of Theorem \ref{Achievability_Theorem}, we invoke Lemma \ref{Logn_lemma} for $\Delta = \sqrt{\frac{\log n}{n}}$.  
\end{remark}

\textit{Proof:} The proof of Lemma \ref{Logn_lemma} is given in Appendix \ref{Logn_lemmaproof}. 

\begin{remark}
The set $\mathcal{P}_n^*$ in both Lemmas \ref{Qccratiolemma} and \ref{Logn_lemma} is a "high probability" set in the following sense. For $\Delta = \omega(n^{-1/2})$, we have $\mathbb{P}\left( \mathbf{Y} \in \mathcal{P}_n^* \right) \to 1$ as $n \to \infty$, if $\mathbb{E}\left [ ||\mathbf{Y}||^2/n \right] = \Gamma + N$ and $\text{Var}(||\mathbf{Y}||^2) = O(n)$. The latter two conditions hold when $\mathbf{Y} \sim Q^*$ or $\mathbf{Y} \sim Q^{cc}$. This property is used multiple times in the proof of Theorem \ref{Achievability_Theorem}.   
\end{remark}

\begin{remark}
    See \cite[Theorem 42]{5452208} and \cite[Proposition 2]{7300429} for similar results to Lemma \ref{Logn_lemma}. 
\end{remark}

\begin{theorem}
    Fix an arbitrary $\epsilon \in (0, 1)$. Consider an AWGN channel with noise variance $N > 0$. Under the mean and variance cost constraint specified by the pair $(\Gamma,V) \in (0, \infty) \times [0, \infty)$, we have 
   \begin{align*}
     \liminf_{n \to \infty}\,\frac{\log M^*(n, \epsilon, \Gamma, V) - nC(\Gamma)}{\sqrt{n}} \geq \max \left \{r \in \mathbb{R} : \mathcal{K}\left(\frac{r}{\sqrt{V(\Gamma) }}, \frac{C'(\Gamma)^2 V}{V(\Gamma)} \right) \leq \epsilon \right \}. 
\end{align*}
Alternatively, for any second-order coding rate $r \in \mathbb{R}$, 
\begin{align*}
    \limsup_{n \to \infty} \bar{P}_{\text{e}}\left(n,C(\Gamma) + \frac{r}{\sqrt{n}},  \Gamma, V\right) &\leq \mathcal{K}\left (\frac{r}{\sqrt{V(\Gamma) }}, \frac{C'(\Gamma)^2 V}{V(\Gamma)}\right ). 
\end{align*}
\label{Achievability_Theorem}
\end{theorem}

\begin{remark}
When $V = 0$, Theorem \ref{Achievability_Theorem} recovers the optimal second-order coding rate corresponding to the maximal cost constraint $||\mathbf{X}|| \leq \sqrt{n\Gamma}$, as given in \cite[Theorem 5]{5290292}. In this special case, the achievability scheme in Theorem \ref{Achievability_Theorem} simplifies to taking the channel input to be uniformly distributed on a singular $(n-1)$-sphere of radius $\sqrt{n\Gamma}$. Thus, for $V = 0$, the proof of Theorem \ref{Achievability_Theorem} is an alternative and more direct proof technique than that of \cite[Theorem 5]{5290292}, which relies on randomly selecting the channel input from sequences of a fixed $n$-type $P^{(n)}$ over a finite alphabet of size $n^{1/4}$ and an implicit assumption on the convergence rate of $I(P^{(n)}, W)$ to $C(\Gamma)$ \cite[p. 4963]{5290292}.
\end{remark}

\begin{remark}
    When $V > 0$, the achievability scheme involves the random channel input having a mixture distribution of two or three uniform distributions on $(n-1)$-spheres.   
\end{remark}

\begin{figure}[H]
    \centering
\includegraphics[width=10cm]{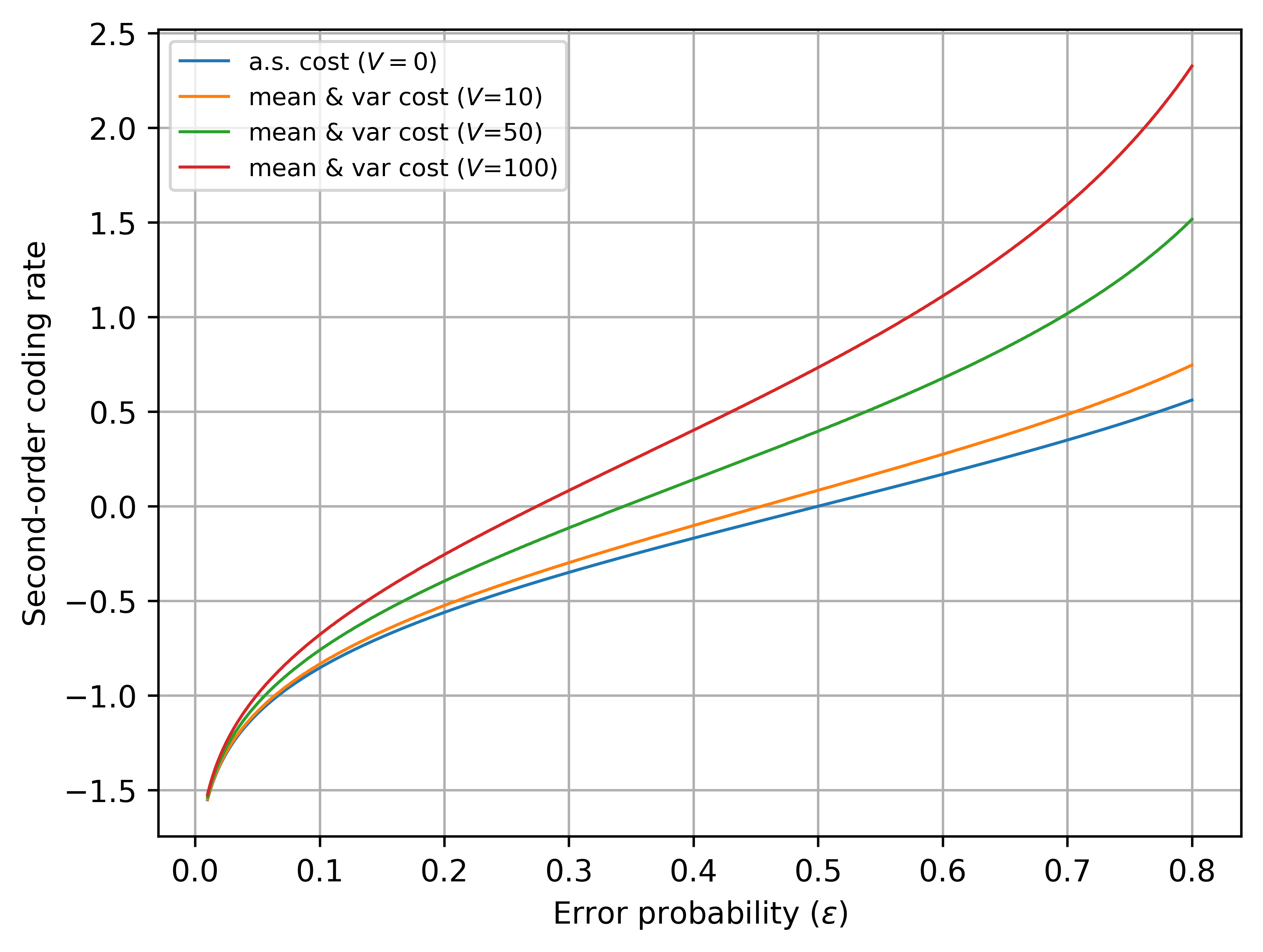}
\caption{The SOCR is compared between the almost-sure cost constraint and the mean and variance cost constraints for different values of $V$. The plots for the mean and variance cost constraints are technically lower bounds to the SOCR since they are obtained through a non-exhaustive search of the feasible region in the maximization and minimization in the formulation of $r(\epsilon, \Gamma, V)$.  }
\label{awgnplots}
\end{figure}

Let $r(\epsilon, \Gamma, V)$ denote the achievable SOCR for the mean and variance cost constraint in Theorem \ref{Achievability_Theorem}, which is also the optimal SOCR as shown later in Theorem \ref{Converse_Theorem}. As remarked earlier, $r(\epsilon, \Gamma, 0) = \sqrt{V(\Gamma)} \Phi^{-1}(\epsilon)$ is the optimal SOCR for the maximal cost constraint. Fig. \ref{awgnplots} plots the SOCR against the average error probability for a Gaussian channel with $\Gamma = 2$ and $N = 1$, showing improved SOCR for several values of $V$. In fact, \cite[Theorem 1]{mahmood2024improvedchannelcodingperformance} shows that $r(\epsilon, \Gamma, V) > \sqrt{V(\Gamma)} \Phi^{-1}(\epsilon)$ for all $V > 0$. Furthermore, $r(\epsilon, \Gamma, V) \to \infty$ as $V \to \infty$ \cite[p. 1509]{10619384}.

\begin{IEEEproof}[Proof of Theorem \ref{Achievability_Theorem}] In view of Lemma \ref{Kfuncproperties}, consider any distribution $P_\Pi$ which achieves the minimum in 
$$\mathcal{K}\left(\frac{r}{\sqrt{V(\Gamma) }}, \frac{C'(\Gamma)^2 V}{V(\Gamma)} \right).$$
Let $\Pi \sim P_\Pi$, where we write 
\begin{align*}
    P_{\Pi}(\pi) = \begin{cases}
    p_1 & \pi = \pi_1\\
    p_2 & \pi = \pi_2\\
    p_3 & \pi = \pi_3.
    \end{cases}
\end{align*}
Recall that $\mathbb{E}[\Pi] = \frac{r}{\sqrt{V(\Gamma) }}$ and $\text{Var}(\Pi) \leq \frac{C'(\Gamma)^2 V}{V(\Gamma)}$. For each $j \in \{1,2,3 \}$, let
\begin{align}
\Gamma_j = \Gamma - \frac{\sqrt{V(\Gamma) }}{C'(\Gamma)\sqrt{n}}\pi_j + \frac{r}{C'(\Gamma)\sqrt{n}}. \label{38n}   
\end{align}
We assume sufficiently large $n$ so that $\Gamma_j > 0$ for all $j \in \{1,2,3 \}$. Let $P^*_j$ be the capacity-cost-achieving input distribution for $C(\Gamma_j)$ and $Q_j^*$ be the corresponding optimal output distribution. Thus, $P_j^* = \mathcal{N}(0, \Gamma_j)$ and $Q_j^* = \mathcal{N}(0, \Gamma_j + N)$. Let $Q^{cc}_j$ be the output distribution induced by the input distribution $\text{Unif}(S^{n-1}_{R_j})$ for $R_j = \sqrt{n \Gamma_j}$. 

\textbf{Achievability Scheme:} Let the random channel input $\mathbf{X}$ be such that with probability $p_j$, $\mathbf{X} \sim \text{Unif}(S^{n-1}_{R_j})$. Denoting the distribution of $\mathbf{X}$ by $P$, we can write 
\begin{align}
    P = \sum_{j=1}^3 p_j \cdot \text{Unif}(S^{n-1}_{R_j}). \label{53vn}
\end{align}
The output distribution of $\mathbf{Y}$ induced by $P \circ W$ is 
\begin{align*}
    PW(\mathbf{y}) &=  \sum_{j=1}^3 p_j Q_j^{cc}(\mathbf{y}).
\end{align*}
We define 
\begin{align*}
    \mathcal{E}_n \coloneqq (P \circ W)\left(\frac{1}{n} \log \frac{W(\mathbf{Y}|\mathbf{X})}{PW(\mathbf{Y})} \leq C(\Gamma) + \frac{r}{\sqrt{n}} \right)
\end{align*}
and show that $\limsup_{n \to \infty} \mathcal{E}_n < \epsilon$ which, by Lemma \ref{aaron'slemma}, would show that the random coding scheme achieves a second-order coding rate of
$r$. 

\textbf{Analysis:}
We first write 
\begin{align}
    \mathcal{E}_n &= \sum_{j=1}^3 p_j \mathbb{P}_{\mathbf{X} \sim \text{Unif}(S^{n-1}_{R_j})} \left(\frac{1}{n} \log \frac{W(\mathbf{Y}|\mathbf{X})}{PW(\mathbf{Y})} \leq C(\Gamma) + \frac{r}{\sqrt{n}} \right).   \label{epsnu} 
\end{align}
To proceed further, we upper bound 
\begin{align}
    &\mathbb{P}_{\mathbf{X} \sim \text{Unif}(S^{n-1}_{R_j})} \left(\frac{1}{n} \log \frac{W(\mathbf{Y}|\mathbf{X})}{PW(\mathbf{Y})} \leq C(\Gamma) + \frac{r}{\sqrt{n}}   \right) \label{referbackn}\\
    &= \int_{\mathbf{y}  \in \mathbb{R}^n}  d \mathbf{y} Q_j^{cc}(\mathbf{y}) \mathbb{P} \left(\frac{1}{n} \log \frac{W(\mathbf{y}|\mathbf{X})}{PW(\mathbf{y})} \leq C(\Gamma) + \frac{r}{\sqrt{n}} \Big | \mathbf{Y} = \mathbf{y} \right) \notag \\
    &\leq \int_{\mathbf{y} \in \mathbb{R}^n } d\mathbf{y} Q_j^{cc}(\mathbf{y}) \mathbb{P} \left(\frac{1}{n} \log \frac{W(\mathbf{y}|\mathbf{X})}{Q_i^{cc}(\mathbf{y})} \leq C(\Gamma) + \frac{r}{\sqrt{n}} \Big |  \mathbf{Y} = \mathbf{y} \right), \label{5cp} 
\end{align}
where $i \in \{1,2,3 \}$ depends on $\mathbf{y}$ and is such that $Q_i^{cc}(\mathbf{y})$ assigns the highest probability to $\mathbf{y}$. We continue the derivation from $(\ref{5cp})$ as 
\begin{align}
&\int_{\mathbf{y} \in \mathbb{R}^n } d\mathbf{y} Q_j^{cc}(\mathbf{y}) \mathbb{P} \left( \log \frac{W(\mathbf{y}|\mathbf{X})}{Q_i^{cc}(\mathbf{y})} \leq n C(\Gamma) + r\sqrt{n} \Big |  \mathbf{Y} = \mathbf{y} \right) \notag \\
    &= \int_{\mathbf{y} \in \mathbb{R}^n } d\mathbf{y} Q_j^{cc}(\mathbf{y}) \mathbb{P} \left( \log \frac{W(\mathbf{y}|\mathbf{X})}{Q_j^{cc}(\mathbf{y})} \leq nC(\Gamma) + r \sqrt{n} + \log \frac{Q_i^{cc}(\mathbf{y})}{Q_j^{cc}(\mathbf{y})} \Big |  \mathbf{Y} = \mathbf{y} \right) \notag \\ 
    &\leq \int_{\mathbf{y} \in \mathbb{R}^n } d\mathbf{y} Q_j^{cc}(\mathbf{y}) \mathbb{P} \left( \log \frac{W(\mathbf{y}|\mathbf{X})}{Q_j^{cc}(\mathbf{y})} \leq nC(\Gamma) + r \sqrt{n} + \kappa_1 \log n \Big |  \mathbf{Y} = \mathbf{y} \right) + \delta_n^{(j)}. \label{1zp}
\end{align}
In the last inequality above,  we used Lemma \ref{Qccratiolemma}. Specifically, in Lemma \ref{Qccratiolemma}, let 
\begin{itemize}
\item $\Gamma = \Gamma_j$,
\item  $\Gamma' = \Gamma_i$,
    \item  $\epsilon = \Gamma_i - \Gamma_j$ so that $\epsilon = O\left(\frac{1}{\sqrt{n}} \right)$, and 
    \item $\Delta = \sqrt{\frac{\log n}{n}}$.
\end{itemize}
Consequently, $\kappa_1$ is a constant from the result of Lemma \ref{Qccratiolemma} and 
\begin{align*}
    \delta_n^{(j)} = Q_j^{cc} \left( \Bigg | \frac{||\mathbf{Y}||^2}{n} - \Gamma_j - N \Bigg | > \Delta \right).
\end{align*}
It can be verified that for $\mathbf{Y} \sim Q_j^{cc}$, $\mathbb{E}\left [ ||\mathbf{Y}||^2 \right] = n \Gamma_j + n N$ and $\text{Var}(||\mathbf{Y}||^2) = 4 n N \Gamma_j  + 2n N^2$. Thus, we have $\delta_n^{(j)} \to 0$ as $n \to \infty$ using Chebyshev inequality.  

Continuing the derivation from $(\ref{1zp})$, we have 
\begin{align}
    &\int_{\mathbf{y} \in \mathbb{R}^n } d\mathbf{y} Q_j^{cc}(\mathbf{y}) \mathbb{P} \left( \log \frac{W(\mathbf{y}|\mathbf{X})}{Q_j^{cc}(\mathbf{y})} \leq nC(\Gamma) + r \sqrt{n} + \kappa_1 \log n \Big |  \mathbf{Y} = \mathbf{y} \right) + \delta_n^{(j)} \notag \\
    &\leq \int_{\mathbf{y} \in \mathbb{R}^n } d\mathbf{y} Q_j^{cc}(\mathbf{y}) \mathbb{P} \left( \log \frac{W(\mathbf{y}|\mathbf{X})}{Q_j^{*}(\mathbf{y})} \leq nC(\Gamma) + r \sqrt{n} + \kappa_1 \log n + \log \frac{Q_j^{cc}(\mathbf{y})}{Q_j^*(\mathbf{y})} \Big |  \mathbf{Y} = \mathbf{y} \right) + \delta_n^{(j)} \notag \\
    &\leq \int_{\mathbf{y} \in \mathbb{R}^n } d\mathbf{y} Q_j^{cc}(\mathbf{y}) \mathbb{P} \left( \log \frac{W(\mathbf{y}|\mathbf{X})}{Q_j^{*}(\mathbf{y})} \leq nC(\Gamma) + r \sqrt{n} + \kappa \log n  \Big |  \mathbf{Y} = \mathbf{y} \right) + 2\delta_n^{(j)}. \label{2x}
\end{align}
In the last inequality above, we used Lemma \ref{Logn_lemma}. Specifically, in Lemma \ref{Logn_lemma}, let $\Gamma = \Gamma_j$ and $\Delta = \sqrt{\frac{\log n}{n}}$. Consequently, $\kappa$ is a suitable bounding constant for both Lemma \ref{Qccratiolemma} and Lemma \ref{Logn_lemma}.

Continuing the derivation from $(\ref{2x})$, we have 
\begin{align*}
    &\int_{\mathbf{y} \in \mathbb{R}^n } d\mathbf{y} Q_j^{cc}(\mathbf{y}) \mathbb{P} \left( \log \frac{W(\mathbf{y}|\mathbf{X})}{Q_j^{*}(\mathbf{y})} \leq nC(\Gamma) + r \sqrt{n} + \kappa \log n  \Big |  \mathbf{Y} = \mathbf{y} \right) + 2\delta_n^{(j)} \notag \\
    &=  \mathbb{P}_{\mathbf{X} \sim \text{Unif}(S^{n-1}_{R_j})} \left( \log \frac{W(\mathbf{Y}|\mathbf{X})}{Q_j^{*}(\mathbf{Y})} \leq nC(\Gamma) + r \sqrt{n} + \kappa \log n  \right) + 2\delta_n^{(j)}\\
    &= \mathbb{P}_{\mathbf{X} \sim \text{Unif}(S^{n-1}_{R_j})} \left( \sum_{m=1}^{n} \log \frac{W(Y_m|X_m)}{Q_j^{*}(Y_m)} - n C(\Gamma_j) \leq n\left( C(\Gamma) - C(\Gamma_j) \right) + r \sqrt{n} + \kappa \log n  \right) + 2\delta_n^{(j)}\\
    &\stackrel{(a)}{=} \mathbb{P}_{\mathbf{X} \sim \text{Unif}(S^{n-1}_{R_j})} \left( \sum_{m=1}^{n}\left [ \log \frac{W(Y_m|X_m)}{Q_j^{*}(Y_m)} - \mathbb{E}\left[ \log \frac{W(Y_m|X_m)}{Q_j^{*}(Y_m)}\right] \right] \leq n\left( C(\Gamma) - C(\Gamma_j) \right) + r \sqrt{n} + \kappa \log n  \right) + 2\delta_n^{(j)}\\
    &\stackrel{(b)}{=} \mathbb{P} \left( \sum_{m=1}^{n} \tilde{T}_m \leq n\left( C(\Gamma) - C(\Gamma_j) \right) + r \sqrt{n} + \kappa \log n  \right) + 2\delta_n^{(j)}\\
    &\stackrel{(c)}{=} \mathbb{P} \left( \frac{1}{\sqrt{n V(\Gamma_j)}} \sum_{m=1}^{n} \tilde{T}_m \leq \sqrt{n} \frac{ C(\Gamma) - C(\Gamma_j) }{\sqrt{V(\Gamma_j)}} + \frac{r}{\sqrt{V(\Gamma_j)}} + \frac{\kappa \log n}{\sqrt{n V(\Gamma_j)}}  \right) + 2\delta_n^{(j)}
\end{align*}
\begin{align}
    &\stackrel{(d)}{\leq} \Phi\left( \sqrt{n} \frac{ C(\Gamma) - C(\Gamma_j) }{\sqrt{V(\Gamma_j)}} + \frac{r}{\sqrt{V(\Gamma_j)}} + \frac{\kappa \log n}{\sqrt{n V(\Gamma_j)}}\right) + \delta_n \notag \\
    &\stackrel{(e)}{\leq} \Phi\left( \frac{\sqrt{n}}{2 \sqrt{V(\Gamma_j)}}\left(\frac{\Gamma - \Gamma_j}{N + \Gamma} \right) + \frac{r}{\sqrt{V(\Gamma_j)}} + \frac{2\kappa \log n}{\sqrt{n V(\Gamma_j)}}\right) + \delta_n. \label{subsintosumof3}
\end{align}

In equality $(a)$ above, we used Lemma \ref{oftusedlemma}. Specifically, use Equation $(\ref{equsentimes})$ in Lemma \ref{oftusedlemma} with $\Gamma = \Gamma_j$. 

In equality $(b)$ above, we used Lemma \ref{sphericalsymm}, where the $\tilde{T}_m$'s are i.i.d. and 
\begin{align}
    \tilde{T}_m = \log \frac{W(Y|\sqrt{\Gamma_j})}{Q^*_j(Y)} - \mathbb{E} \left [\log  \frac{W(Y|\sqrt{\Gamma_j})}{Q^*_j(Y)} \right ],     
\end{align}
where $Y \sim \mathcal{N}(\sqrt{\Gamma_j}, N)$. 

In equality $(c)$ above, we normalize the sum to have unit variance, which follows from Corollary \ref{lem2cor} and Equation $(\ref{VGammadef})$. 

In inequality $(d)$ above, we used the Berry-Esseen Theorem \cite{esseen11} to obtain convergence of the CDF of the normalized sum of i.i.d. random variables $\tilde{T}_m$'s to the standard normal CDF, with $\delta_n \to 0$ accounting for both the rate of convergence and $\delta_n^{(j)} \to 0$. In inequality $(e)$ above, we used a Taylor series approximation, noting that $\Gamma - \Gamma_j = O(1/\sqrt{n})$.

We can now upper bound $(\ref{referbackn})$ by $(\ref{subsintosumof3})$, which allows us to upper bound $(\ref{epsnu})$ as 
\begin{align*}
    \mathcal{E}_n &\leq \sum_{j=1}^3 p_j \Phi\left( \frac{\sqrt{n}}{2 \sqrt{V(\Gamma_j)}}\left(\frac{\Gamma - \Gamma_j}{N + \Gamma} \right) + \frac{r}{\sqrt{V(\Gamma_j)}} + \frac{2\kappa \log n}{\sqrt{n V(\Gamma_j)}}\right) + \delta_n
\end{align*}
for some redefined sequence $\delta_n \to 0$ as $n \to \infty$. Using Equation $(\ref{38n})$ and the formula for $C'(\Gamma)$ from Lemma \ref{oftusedlemma}, we can simplify the upper bound as 
\begin{align*}
     \mathcal{E}_n &\leq \sum_{j=1}^3 p_j \Phi\left( \frac{\sqrt{V(\Gamma)}}{\sqrt{V(\Gamma_j)}} \pi_j - \frac{r}{\sqrt{V(\Gamma_j)}} + \frac{r}{\sqrt{V(\Gamma_j)}} + \frac{2\kappa \log n}{\sqrt{n V(\Gamma_j)}}  \right) + \delta_n. 
\end{align*}
Therefore, since $\Gamma_j \to \Gamma$ as $n \to \infty$, we have 
\begin{align}
    \limsup_{n \to \infty} \mathcal{E}_n &\leq \sum_{j=1}^3 p_j \Phi\left( \pi_j \right) \notag \\
    &= \mathcal{K}\left(\frac{r}{\sqrt{V(\Gamma) }}, \frac{C'(\Gamma)^2 V}{V(\Gamma)} \right) \label{extendtoerror}. 
\end{align}
To complete the proof, we choose $r < r^*$ where 
\begin{align}
    r^* = \max \left \{r : \mathcal{K}\left(\frac{r}{\sqrt{V(\Gamma) }}, \frac{C'(\Gamma)^2 V}{V(\Gamma)} \right) \leq \epsilon \right \}. \label{rstar9}
\end{align}
Hence, $\limsup_{n \to \infty} \mathcal{E}_n < \epsilon$ because, from Lemma \ref{Kfuncproperties},
$\mathcal{K}(\cdot,\cdot)$ is a strictly increasing function in the first argument. Hence, $\limsup_{n \to \infty} \mathcal{E}_n < \epsilon$ for $r < r^*$, thus establishing that any $r < r^*$ is an achievable second-order coding rate. Finally, letting $r \to r^*$ establishes an achievable second-order coding rate of $r^*$, matching the converse in Theorem \ref{Converse_Theorem}.

The achievability result can also be stated in terms of an upper bound on the minimum average probability of error of $(n, R, \Gamma, V)$ codes for a rate $R = C(\Gamma) + \frac{r}{\sqrt{n}}$. From Lemma \ref{aaron'slemma}, we have for $\theta = 1/n^{\vartheta}$ for $1 > \vartheta > 1/2$,
\begin{align}
    \bar{P}_{\text{e}}\left(n,C(\Gamma) + \frac{r}{\sqrt{n}},  \Gamma, V\right) \leq (P \circ W) \left(\frac{1}{n} \log \frac{W(\mathbf{Y}|\mathbf{X})}{PW(\mathbf{Y})} \leq C(\Gamma) + \frac{r}{\sqrt{n}} + \frac{1}{n^\vartheta} \right) + e^{-n^{1-\vartheta}}.  \label{notplaying}
\end{align}
For any $r' > r$, we have $\frac{r}{\sqrt{n}} + \frac{1}{n^\vartheta} < \frac{r'}{\sqrt{n}}$ eventually, so
\begin{align}
    &\limsup_{n \to \infty} \bar{P}_{\text{e}}\left(n,C(\Gamma) + \frac{r}{\sqrt{n}},  \Gamma, V\right) \notag  \\
    &\leq \limsup_{n \to \infty} (P \circ W) \left(\frac{1}{n} \log \frac{W(\mathbf{Y}|\mathbf{X})}{PW(\mathbf{Y})} \leq C(\Gamma) + \frac{r'}{\sqrt{n}} \right) \notag\\
    &\leq \mathcal{K}\left (\frac{r'}{\sqrt{V(\Gamma) }}, \frac{C'(\Gamma)^2 V}{V(\Gamma)}\right ), \label{notplayggg}
\end{align}
where the last inequality follows from $(\ref{extendtoerror})$. Letting $r' \to r$ in $(\ref{notplayggg})$ and invoking continuity of the function $\mathcal{K}$ establishes the result.

\end{IEEEproof}

\section{Converse Result \label{main_converse_section}}

\begin{lemma}
Consider an AWGN channel $W$ with noise variance $N > 0$ and cost constraint $(\Gamma,V) \in (0, \infty) \times (0, \infty)$. Then for every $n, \rho > 0$ and $\epsilon \in (0, 1)$,   
\begin{align}
    \log M^*(n, \epsilon, \Gamma, V) &\leq \log \rho - \log \left [ \left( 1 - \epsilon - \sup_{\overline{P} \in \mathcal{P}_{\Gamma,V}(\mathbb{R}^n)}\, \inf_{q \in \mathcal{P}(\mathbb{R}^n)} (\overline{P} \circ W) \left( \frac{W(\mathbf{Y}|\mathbf{X})}{q(\mathbf{Y})} > \rho \right)\right)^+\right], \label{qandp}
\end{align}
where $\mathcal{P}_{\Gamma, V}(\mathbb{R}^n) \subset \mathcal{P}(\mathbb{R}^n)$ is the set of distributions $\overline{P}$ such that $\mathbb{E}\left [ \sum_{i=1}^n c(X_i) \right] \leq n \Gamma$ and $\text{Var}\left(\sum_{i=1}^n c(X_i) \right) \leq n V$ for $\mathbf{X} \sim \overline{P}$.
\label{mostgenconv}
\end{lemma}
\textit{Proof:} The proof of Lemma \ref{mostgenconv} is similar to that of \cite[Lemma 2]{mahmood2025channelcodingmeanvariance} and is omitted.

\begin{remark}
    Lemma \ref{mostgenconv} serves as our starting point for the converse. It is a specialization of Polyanskiy–Poor–Verdú’s meta-converse theorem \cite[Theorem 27]{5452208}. Specifically, we restrict the set of input distributions in \cite[Theorem 27]{5452208} to $\mathcal{P}_{\Gamma, V}(\mathbb{R}^n)$ and then apply a Neyman-Pearson threshold inequality (see, e.g., \cite[(102)]{5452208}): 
\begin{align}
    \left(1 - \epsilon - (\overline{P} \circ W) \left( \frac{W(\mathbf{Y}|\mathbf{X})}{q(\mathbf{Y})} > \rho \right)\right)^+ \leq \rho \,\beta_{1-\epsilon}\left( \overline{P} \circ W, \overline{P} \circ q \right).
\end{align}
\end{remark}

\begin{theorem}
    Fix an arbitrary $\epsilon \in (0, 1)$. Consider an AWGN channel with noise variance $N > 0$. Under the mean and variance cost constraints specified by the pair $(\Gamma,V) \in (0, \infty) \times [0, \infty)$, we have 
   \begin{align}
     \limsup_{n \to \infty}\,\frac{\log M^*(n, \epsilon, \Gamma, V) - nC(\Gamma)}{\sqrt{n}} \leq \max \left \{r \in \mathbb{R} : \mathcal{K}\left(\frac{r}{\sqrt{V(\Gamma) }}, \frac{C'(\Gamma)^2 V}{V(\Gamma)} \right) \leq \epsilon \right \}. \label{upper_bound_V=0}
\end{align}
Alternatively, for any second-order coding rate $r \in \mathbb{R}$, 
\begin{align}
    \liminf_{n \to \infty} \bar{P}_{\text{e}}\left(n,C(\Gamma) + \frac{r}{\sqrt{n}},  \Gamma, V\right) &\geq \mathcal{K}\left (\frac{r}{\sqrt{V(\Gamma) }}, \frac{C'(\Gamma)^2 V}{V(\Gamma)}\right ).  \label{lower_bound_V=0}
\end{align}
    \label{Converse_Theorem}
\end{theorem}

\begin{IEEEproof}[Proof of Theorem \ref{Converse_Theorem}]
    
For $V = 0$, we are required to prove that the upper bound in $(\ref{upper_bound_V=0})$ is $\sqrt{V(\Gamma)} \Phi^{-1}(\epsilon)$, and the lower bound in $(\ref{lower_bound_V=0})$ is $\Phi\left(r/\sqrt{V(\Gamma)} \right)$. But this follows from the known converse result \cite[Theorem 5]{5290292} for the maximal cost constraint, since the m.v. cost constraint for $V = 0$ is more stringent.

We assume $V > 0$ for the remainder of the proof. We start with Lemma \ref{mostgenconv} and first upper bound
\begin{align}
    \sup_{\overline{P} \in \mathcal{P}_{\Gamma,V}(\mathbb{R}^n)}\, \inf_{q \in \mathcal{P}(\mathbb{R}^n)} (\overline{P} \circ W) \left( \frac{W(\mathbf{Y}|\mathbf{X})}{q(\mathbf{Y})} > \rho \right). \label{bx}
\end{align}

Let $\rho = \exp\left(n C(\Gamma) + \sqrt{n} r \right)$ in $(\ref{bx})$, where $r$ is a number to be specified later. Let $\mathbf{X} \sim \overline{P}$ for any arbitrary $\overline{P} \in \mathcal{P}_{\Gamma,V}(\mathbb{R}^n)$ so that $\mathbb{E}\left [ \sum_{i=1}^n c(X_i) \right ] \leq n \Gamma$ and $\text{Var}\left(\sum_{i=1}^n c(X_i) \right) \leq n V$. Choosing $q = Q^*$ in $(\ref{bx})$, we have 
\begin{align}
    &(\overline{P} \circ W) \left( \log \frac{W(\mathbf{Y}|\mathbf{X})}{Q^*(\mathbf{Y})} > n C(\Gamma) + \sqrt{n} r  \right) \notag \\
    &= \int_{\mathbb{R}^n} d \overline{P}(\mathbf{x}) \mathbb{P}_{W(\cdot|\mathbf{x})}\left( \sum_{i=1}^n \log \frac{W(Y_i|x_i)}{Q^*(Y_i)} > n C(\Gamma) + \sqrt{n} r  \right), \label{fopv}
\end{align}
where in $(\ref{fopv})$, $Y_1, Y_2, \ldots, Y_n$ are independent random variables and each $Y_i \sim \mathcal{N}(x_i, N)$. We will now proceed with upper bounding 
$$\mathbb{P}_{W(\cdot|\mathbf{x})}\left( \sum_{i=1}^n \log \frac{W(Y_i|x_i)}{Q^*(Y_i)} > n C(\Gamma) + \sqrt{n} r  \right)$$
for different sets of input vectors $\mathbf{x}$ based on their empirical distributions $P_{\mathbf{x}}$. We define the empirical distribution $P_{\mathbf{x}}$ of a vector $\mathbf{x}$ as  
\begin{align}
    P_{\mathbf{x}} = \frac{1}{n} \sum_{i=1}^n \delta_{x_i},
\end{align}
where $\delta_x$ is the dirac delta measure at $x$. For any given channel input $\mathbf{x}$ and independent $Y_i$'s where $Y_i \sim \mathcal{N}(x_i, N)$, we have from Corollary \ref{lem2cor} that
\begin{align*}
    &\text{Var}\left( \frac{1}{\sqrt{n}} \sum_{i=1}^n \left [   \log \frac{W(Y_{i}|x_i)}{Q^*(Y_i)}  - \mathbb{E} \left [   \log \frac{W(Y_i|x_i)}{Q^*(Y_i)}  \right]\right] \right)\\
    &= \int \nu_{t} P_{\mathbf{x}}(t) dt\\ 
 &= \frac{\Gamma^2 + 2N c(\mathbf{x})}{2(N + \Gamma)^2}
\end{align*}

For any $\beta \geq 0$, define 
\begin{align}
    \tau_\beta \coloneqq \sup_{\Gamma - \beta < t \leq \Gamma + \beta} \Bigg |\frac{\Gamma^2 + 2N t}{2(N + \Gamma)^2} - V(\Gamma) \Bigg | \label{kof} 
\end{align}
and note that $\tau_\beta \to 0$ as $\beta \to 0$. Define the function 
\begin{align}
    \psi_\beta\left( t \right ) \coloneqq \begin{cases}
V(\Gamma) + \tau_\beta & \text{ if } t \leq \Gamma\\
V(\Gamma) - \tau_\beta & \text{ if } t > \Gamma.
\end{cases}    
\end{align}

Fix any $0 < \Delta < 1/2$ and $\eta > 0$. Then select $\beta > 0$ such that $\tau_\beta < V(\Gamma)$, 
\begin{align}
    \sup_{t \in \mathbb{R}} \Bigg | \frac{1}{\sqrt{V(\Gamma)}} - \frac{1}{\sqrt{\psi_\beta(t)}} \Bigg | \leq \frac{ \Delta}{C'(\Gamma) \sqrt{V}}\label{aaronerr1}
\end{align}
and 
\begin{align}
    \sup_{t \in \mathbb{R}} \Bigg | \frac{1}{V(\Gamma)} - \frac{1}{\psi_\beta(t)} \Bigg | \leq \frac{ \Delta/2}{C'(\Gamma)^2 V}. \label{aaronerr2}
\end{align}

Also define 
\begin{align}
    \varphi(r) \coloneqq \begin{cases}
V(\Gamma) - \tau_\beta & \text{ if } r \leq 0\\
V(\Gamma) + \tau_\beta & \text{ if } r > 0.
\end{cases} \label{varphidefinitionr}
\end{align} 
With $\Delta$, $\eta$ and $\beta$ fixed as above\footnote{$\eta$ is used later.}, we divide the set of sequences $\mathbf{x} \in \mathbb{R}^n$ into three subsets:
\begin{align}
\begin{split}
    \mathcal{P}_{n,1} &\coloneqq \left \{ \mathbf{x}: c(\mathbf{x}) \leq \Gamma - \beta \right \},\\
    \mathcal{P}_{n,2} &\coloneqq \left \{ \mathbf{x}: \Gamma - \beta < c(\mathbf{x}) \leq \Gamma + \beta \right \},\\
    \mathcal{P}_{n,3} &\coloneqq \left \{ \mathbf{x}: c(\mathbf{x}) > \Gamma + \beta \right \}.
    \end{split}
\end{align}
With the choice of $\rho$ as $\rho = \exp \left(n C(\Gamma) + \sqrt{n} r \right)$, we seek to establish a first-order converse rate of $C(\Gamma)$ and a second-order converse rate of $r/\sqrt{n}$. Over the set $\mathcal{P}_{n,1}$, standard concentration arguments suffice since the cost $c(\mathbf{x})$ being bounded away from $\Gamma$ implies that the mean of the information density will be bounded away from $C(\Gamma)$. For $\mathcal{P}_{n,3}$, the variance constraint will make the probability of the set negligible. For $\mathcal{P}_{n,2}$, we will do central limit theorem-based analysis. The conditions in $(\ref{kof})-(\ref{varphidefinitionr})$ will allow us to uniformly bound the variance of the information density 
over all $\mathbf{x} \in \mathcal{P}_{n,2}$, which is important in the application of the Berry-Esseen Theorem \cite{esseen11}.

For $\mathbf{x} \in \mathcal{P}_{n, 1}$, we have 
\begin{align}
  &\mathbb{P}_{W(\cdot|\mathbf{x})} \left( \sum_{i=1}^n  \log \frac{W(Y_i|x_i)}{Q^*(Y_i)} > n C(\Gamma) + \sqrt{n} r \right) \notag \\
    &\stackrel{(a)}{=} \mathbb{P}_{W(\cdot|\mathbf{x})} \left( \sum_{i=1}^n \left [   \log \frac{W(Y_{i}|x_i)}{Q^*(Y_i)}  
 - \mathbb{E} \left [   \log \frac{W(Y_i|x_i)}{Q^*(Y_i)}  \right]\right] >     nC'(\Gamma)\left (\Gamma - c(\mathbf{x})\right )  + \sqrt{n} r  
 \right) \notag \\
&\leq \mathbb{P}_{W(\cdot|\mathbf{x})} \left( \sum_{i=1}^n  \left [   \log \frac{W(Y_{i}|x_i)}{Q^*(Y_i)}  
 - \mathbb{E} \left [   \log \frac{W(Y_i|x_i)}{Q^*(Y_i)}  \right]\right] >     nC'(\Gamma)\beta  + \sqrt{n} r  \right) \notag \\
 &\stackrel{(b)}{\leq} \mathbb{P}_{W(\cdot|\mathbf{x})} \left( \sum_{i=1}^n \left [   \log \frac{W(Y_{i}|x_i)}{Q^*(Y_i)}  
 - \mathbb{E} \left [   \log \frac{W(Y_i|x_i)}{Q^*(Y_i)}  \right]\right] >     \frac{nC'(\Gamma)\beta}{2}    \right) \notag \\
 &\stackrel{(c)}{\leq} \frac{4}{n C'(\Gamma)^2 \beta^2} \left( \frac{\Gamma^2 + 2N \Gamma}{2(N + \Gamma)^2} \right)  \label{pn1}
\end{align}
for sufficiently large $n$. In equality $(a)$, we used Lemma \ref{oftusedlemma}. Inequality $(b)$ holds because $r$ is a constant. Inequality $(c)$ follows by applying Chebyshev's inequality and Corollary \ref{lem2cor}, the latter of which gives us that
\begin{align*}
    &\text{Var}\left( \sum_{i=1}^n \left [   \log \frac{W(Y_{i}|x_i)}{Q^*(Y_i)}  
 - \mathbb{E} \left [   \log \frac{W(Y_i|x_i)}{Q^*(Y_i)}  \right]\right] \right)\\
 &= \frac{n\Gamma^2 + 2nN c(\mathbf{x})}{2\left(N + \Gamma \right)^2}\\
 &\leq n \left( \frac{\Gamma^2 + 2N \Gamma}{2(N + \Gamma)^2} \right). 
\end{align*}
For $\mathbf{x} \in \mathcal{P}_{n,2}$, we have 
\begin{align}
    & \mathbb{P}_{W(\cdot|\mathbf{x})} \left( \sum_{i=1}^n  \log \frac{W(Y_i|x_i)}{Q^*(Y_i)} > n C(\Gamma) + \sqrt{n} r  \right) \notag \\
    &\stackrel{(a)}{=} \mathbb{P}_{W(\cdot|\mathbf{x})} \left( \sum_{i=1}^n \left [   \log \frac{W(Y_{i}|x_i)}{Q^*(Y_i)}  
 - \mathbb{E} \left [   \log \frac{W(Y_i|x_i)}{Q^*(Y_i)}  \right]\right] >     nC'(\Gamma)\left (\Gamma - c(\mathbf{x})\right )  + \sqrt{n} r  
 \right) \notag \\
 &\stackrel{(b)}{=} \mathbb{P}_{W(\cdot|\mathbf{x})} \left( \frac{1}{\sqrt{\int n\,\nu_t P_{\mathbf{x}}(t) dt}} \sum_{i=1}^n \left [   \log \frac{W(Y_{i}|x_i)}{Q^*(Y_i)}  
 - \mathbb{E} \left [   \log \frac{W(Y_i|x_i)}{Q^*(Y_i)}  \right]\right] > \mbox{} \right. \notag \\
 & \left. \quad \quad \quad \quad \quad \quad \quad \quad \quad \frac{ \sqrt{n}C'(\Gamma)\left (\Gamma - c(\mathbf{x})\right )}{\sqrt{\int \nu_t P_{\mathbf{x}}(t) dt}}  + \frac{ r}{\sqrt{\int \nu_t P_{\mathbf{x}}(t) dt}}  \right) \notag \\
 &\stackrel{(c)}{=} \mathbb{P}_{W(\cdot|\mathbf{x})} \left( \frac{1}{\sqrt{\int n\,\nu_t P_{\mathbf{x}}(t) dt}} \sum_{i=1}^n \tilde{T}_i  > \frac{ \sqrt{n}C'(\Gamma)\left (\Gamma - c(\mathbf{x})\right )}{\sqrt{\int \nu_t P_{\mathbf{x}}(t) dt}}  + \frac{ r}{\sqrt{\int \nu_t P_{\mathbf{x}}(t) dt}}  \right)  \notag\\ 
 &\stackrel{(d)}{\leq} 1 - \Phi \left( \frac{ \sqrt{n}C'(\Gamma)\left (\Gamma - c(\mathbf{x})\right )}{\sqrt{\int \nu_t P_{\mathbf{x}}(t) dt}}  + \frac{ r}{\sqrt{\int \nu_t P_{\mathbf{x}}(t) dt}}  
 \right) + \frac{\kappa_{\mathbf{x}}}{\sqrt{n}}\notag \\
 &\stackrel{(e)}{\leq} 1 - \Phi \left( \frac{ \sqrt{n}C'(\Gamma)\left (\Gamma - c(\mathbf{x})\right )}{\sqrt{\int \nu_t P_{\mathbf{x}}(t) dt}}  + \frac{ r}{\sqrt{\int \nu_t P_{\mathbf{x}}(t) dt}}  
 \right) + \frac{\kappa}{\sqrt{n}} \notag  \\
 &\leq 1 - \Phi \left( \frac{ \sqrt{n}C'(\Gamma)\left (\Gamma - c(\mathbf{x})\right )}{\sqrt{\psi_\beta(c(\mathbf{x}))}}  + \frac{ r}{\sqrt{\varphi(r)}}  
 \right) + \frac{\kappa}{\sqrt{n}}   \label{pn2}
\end{align}
for sufficiently large $n$. In equality $(a)$, we used Lemma \ref{oftusedlemma}. In equality $(b)$, we normalize the sum to have unit variance, where we write 
\begin{align*}
   &\int_{-\infty}^\infty n\,\nu_t P_{\mathbf{x}}(t) dt\\
   &= \text{Var}\left( \sum_{i=1}^n \left [   \log \frac{W(Y_{i}|x_i)}{Q^*(Y_i)}  
 - \mathbb{E} \left [   \log \frac{W(Y_i|x_i)}{Q^*(Y_i)}  \right]\right] \right)\\
 &=  \frac{n\Gamma^2 + 2nN c(\mathbf{x})}{2\left(N + \Gamma \right)^2}.
\end{align*}
In equality $(c)$, we use Lemma \ref{sphericalsymm}.
In inequality $(d)$, we apply the Berry-Esseen Theorem \cite{esseen11}, where $\kappa_{\mathbf{x}}$ is a constant depending on the second- and third-order moments of $\{\tilde{T}_i \}_{i=1}^n$.
In inequality $(e)$, we use the fact that the distribution of each $\tilde{T}_i$ depends on $\mathbf{x}$ only through its cost $c(\mathbf{x})$. Since $c(\mathbf{x})$ is uniformly bounded over the set $\mathcal{P}_{n, 2}$, the constant $\kappa_{\mathbf{x}}$ can be uniformly upper bounded over all $\mathbf{x} \in \mathcal{P}_{n, 2}$ by some constant $\kappa$ that does not depend on $\mathbf{x}$ at all.

For $\mathcal{P}_{n, 3}$, the following lemma shows that the probability of this set under $\overline{P}$ is small.

\begin{lemma}
    We have 
    \begin{align}
\overline{P}\left(\mathcal{P}_{n,3} \right) \leq \frac{V}{n \beta^2}. \label{pn3}    
    \end{align}
    \end{lemma}
\begin{IEEEproof}
Let $\mu_i = \mathbb{E}[c(X_i)]$. For any $\beta > 0$, 
\begin{align*}
    &\overline{P}\left( c(\mathbf{X}) > \Gamma + \beta \right)\\
    &= \overline{P}\left( \sum_{i=1}^n \left (c(X_i) - \mu_i\right ) + \sum_{i=1}^n \mu_i \geq n\Gamma + n\beta \right)\\
    &\leq  \overline{P}\left( \sum_{i=1}^n \left (c(X_i) - \mu_i\right ) \geq  n\beta \right)\\
    &\leq \overline{P}\left( \left( \sum_{i=1}^n \left (c(X_i) - \mu_i\right )\right)^2 \geq  n^2\beta^2 \right)\\
    &\leq \frac{1}{n^2 \beta^2} \mathbb{E} \left [ \left( \sum_{i=1}^n \left (c(X_i) - \mu_i\right )\right)^2 \right ]\\
    &\leq \frac{V}{n \beta^2}.  
\end{align*}

\end{IEEEproof}

Using the results in $(\ref{pn1})$, $(\ref{pn2})$ and $(\ref{pn3})$, we can upper bound $(\ref{fopv})$ as 
\begin{align}
    &(\overline{P} \circ W) \left( \log \frac{W(\mathbf{Y}|\mathbf{X})}{Q^*(\mathbf{Y})} > nC(\Gamma) + \sqrt{n} r \right) \notag \\
    &\leq 1 + \frac{4}{n C'(\Gamma)^2 \beta^2} \left( \frac{\Gamma^2 + 2N \Gamma}{2(N + \Gamma)^2} \right) + \frac{V}{n \beta^2} + \frac{\kappa}{\sqrt{n}}  - \mbox{} \notag \\
    & \quad \quad \quad \mathbb{E}\left[ \Phi \left( \frac{ \sqrt{n}C'(\Gamma)\left (\Gamma - c(\mathbf{X})\right )}{\sqrt{\psi_\beta(c(\mathbf{X}))}}  + \frac{ r}{\sqrt{\varphi(r)}}  
 \right) \right ].  \label{pq}
\end{align}
To further upper bound the above expression, we need to obtain a lower bound to 
\begin{align}
    \inf_{\mathbf{X}}  \, \mathbb{E}\left[ \Phi \left( \frac{ \sqrt{n}C'(\Gamma)\left (\Gamma - c(\mathbf{X})\right )}{\sqrt{\psi_\beta(c(\mathbf{X}))}}  + \frac{ r}{\sqrt{\varphi(r)}}  
 \right) \right ],
\end{align}
where the infimum is over all random vectors $\mathbf{X}$ such that $\mathbb{E}\left[ c(\mathbf{X}) \right] \leq \Gamma$ and $\text{Var}\left(c(\mathbf{X}) \right) \leq V/n$. Without loss of generality, we can assume $\mathbb{E}[c(\mathbf{X})] = \Gamma$ since the function $\Phi$ is monotonically increasing and the function 
$$c \mapsto \frac{\Gamma - c}{\sqrt{\psi_\beta(c)}}$$
is monotonically nonincreasing.  

Note that  
\begin{align}
    &\Bigg | \mathbb{E}\left [ \frac{\sqrt{n} C'(\Gamma)}{\sqrt{\psi_{\beta}(c(\mathbf{X}))}} \left(\Gamma - c(\mathbf{X}) \right) \right] - \mathbb{E}\left [ \frac{\sqrt{n} C'(\Gamma)}{\sqrt{V(\Gamma)}} \left(\Gamma - c(\mathbf{X}) \right) \right] \Bigg | \notag \\
    &= \Bigg | \mathbb{E}\left [ \sqrt{n} C'(\Gamma)  \left(\Gamma - c(\mathbf{X}) \right)\left [ \frac{1}{\sqrt{\psi_{\beta}(c(\mathbf{X}))}} - \frac{1}{\sqrt{V(\Gamma)}} \right ] \right] \Bigg | \notag \\
    &\leq  \sqrt{n} C'(\Gamma) \mathbb{E} \left [ \big | \Gamma - c(\mathbf{X}) \big | \cdot \Bigg |  \frac{1}{\sqrt{\psi_{\beta}(c(\mathbf{X}))}} - \frac{1}{\sqrt{V(\Gamma)}} \Bigg | \right ] \notag \\
    &\stackrel{(a)}{\leq} \sqrt{n} C'(\Gamma) \frac{ \Delta}{C'(\Gamma) \sqrt{V}} \mathbb{E}\left [ \big |\Gamma -  c(\mathbf{X}) \big | \right] \notag \\
    &\stackrel{(b)}{\leq} \Delta. \label{bzuyv} 
\end{align}
In inequality $(a)$, we used $(\ref{aaronerr1})$. In inequality $(b)$, we used the fact that $\mathbb{E}\left [ \sqrt{ \left ( \Gamma - c(\mathbf{X})  \right)^2 }\right] \leq \sqrt{\text{Var}(c(\mathbf{X}))}$. Hence, from $(\ref{bzuyv})$, we have 
\begin{align}
    &\mathbb{E}\left [ \frac{\sqrt{n} C'(\Gamma)}{\sqrt{\psi_{\beta}(c(\mathbf{X}))}} \left(\Gamma - c(\mathbf{X}) \right) \right] \notag \\
    &\geq \mathbb{E}\left [ \frac{\sqrt{n} C'(\Gamma)}{\sqrt{V(\Gamma)}} \left(\Gamma - c(\mathbf{X}) \right) \right] - \Delta \notag \\
    &= -\Delta. \label{Piprimemean}
\end{align}

Now define $X' = \Gamma - c(\mathbf{X})$ so that $\mathbb{E}[X'] = 0$ and $\text{Var}(X') \leq V/n$. Also define 
\begin{align*}
    \Psi_\beta( X' ) \coloneqq \begin{cases}
V(\Gamma) + \tau_\beta & \text{ if } X' \geq 0\\
V(\Gamma) - \tau_\beta & \text{ if } X' < 0.
\end{cases}    
\end{align*}
Then note that  
\begin{align}
    & \Bigg |\text{Var}\left( \frac{\Gamma - c(\mathbf{X})}{\sqrt{\psi_\beta(c(\mathbf{X}))}} \right) -
     \text{Var}\left( \frac{\Gamma - c(\mathbf{X})}{\sqrt{V(\Gamma)}} \right) \Bigg | \notag  \\
     &= \Bigg | \text{Var}\left( \frac{X'}{\sqrt{\Psi_\beta(X')}} \right) - \text{Var}\left( \frac{X'}{\sqrt{V(\Gamma)}} \right) \Bigg | \notag \\
     &= \Bigg | \mathbb{E}\left [ \frac{X'^2}{\Psi_\beta(X')}  - \frac{X'^2}{V(\Gamma)} \right ] +  \left(\mathbb{E}\left [  \frac{X'}{\sqrt{V(\Gamma)}} \right] \right)^2 - \left(\mathbb{E}\left [  \frac{X'}{\sqrt{\Psi_\beta(X')}} \right] \right)^2 \Bigg |\notag  \\
     &=  \Bigg | \mathbb{E}\left [ \left( \frac{X'}{\sqrt{\Psi_\beta(X')}} + \frac{X'}{\sqrt{V(\Gamma)}}\right) \left( \frac{X'}{\sqrt{\Psi_\beta(X')}}  -\frac{X'}{\sqrt{V(\Gamma)}}\right)   \right ] +  \mbox{} \notag \\
     & \quad \quad \quad \quad \quad \left(\mathbb{E}\left [  \frac{X'}{\sqrt{V(\Gamma)}} \right] + \mathbb{E}\left [  \frac{X'}{\sqrt{\Psi_\beta(X')}} \right] \right) \left (\mathbb{E}\left [  \frac{X'}{\sqrt{V(\Gamma)}} \right] - \mathbb{E}\left [  \frac{X'}{\sqrt{\Psi_\beta(X')}} \right] \right) \Bigg | \notag \\
     &\leq \Bigg | \mathbb{E}\left [ \left( \frac{X'}{\sqrt{\Psi_\beta(X')}} + \frac{X'}{\sqrt{V(\Gamma)}}\right) \left( \frac{X'}{\sqrt{\Psi_\beta(X')}}  -\frac{X'}{\sqrt{V(\Gamma)}}\right)   \right ] \Bigg | +  \mbox{} \notag  \\
     & \quad \quad \quad \quad \quad \Bigg |\left(\mathbb{E}\left [  \frac{X'}{\sqrt{V(\Gamma)}} \right] + \mathbb{E}\left [  \frac{X'}{\sqrt{\Psi_\beta(X')}} \right] \right) \left (\mathbb{E}\left [  \frac{X'}{\sqrt{V(\Gamma)}} \right] - \mathbb{E}\left [  \frac{X'}{\sqrt{\Psi_\beta(X')}} \right] \right) \Bigg | \notag \\
     &\stackrel{(a)}{\leq} \mathbb{E}[X'^2] \frac{ \Delta/2}{C'(\Gamma)^2 V} + \Bigg |\left(\mathbb{E}\left [  \frac{X'}{\sqrt{V(\Gamma)}} \right] + \mathbb{E}\left [  \frac{X'}{\sqrt{\Psi_\beta(X')}} \right] \right) \left (\mathbb{E}\left [  \frac{X'}{\sqrt{V(\Gamma)}} \right] - \mathbb{E}\left [  \frac{X'}{\sqrt{\Psi_\beta(X')}} \right] \right) \Bigg | \notag \\
     &\leq \mathbb{E}[X'^2] \frac{ \Delta/2}{C'(\Gamma)^2 V} + \Bigg |\mathbb{E}\left [  \frac{X'}{\sqrt{V(\Gamma)}} +  \frac{X'}{\sqrt{\Psi_\beta(X')}} \right] \Bigg | \cdot \Bigg |\mathbb{E}\left [  \frac{X'}{\sqrt{V(\Gamma)}}  -  \frac{X'}{\sqrt{\Psi_\beta(X')}} \right] \Bigg | \notag \\
     &\leq \mathbb{E}[X'^2] \frac{ \Delta/2}{C'(\Gamma)^2 V} + \Bigg | \mathbb{E}\left [   \frac{X'}{\sqrt{\Psi_\beta(X')}}   \right] \Bigg | \cdot \mathbb{E}\left [ |X'| \cdot \Bigg | \frac{1}{\sqrt{V(\Gamma)}} -  \frac{1}{\sqrt{\Psi_\beta(X')}} \Bigg | \right] \notag  \\
     &\stackrel{(b)}{\leq} \mathbb{E}[X'^2] \frac{ \Delta/2}{C'(\Gamma)^2 V} + \frac{\Delta}{\sqrt{n} C'(\Gamma)}  \cdot  \frac{ \Delta}{C'(\Gamma) \sqrt{V}}  \mathbb{E}\left [ |X'|   \right]  \notag \\
     &\stackrel{(c)}{\leq} \frac{V}{n} \frac{ \Delta/2}{C'(\Gamma)^2 V} + \frac{\Delta}{\sqrt{n} C'(\Gamma)}  \cdot  \frac{ \Delta}{C'(\Gamma) \sqrt{V}}  \sqrt{\frac{V}{n}} \notag \\
     &=  \frac{1}{n C'(\Gamma)^2} \left( \frac{ \Delta}{2} + \Delta^2 \right)\notag \\
     &\stackrel{(d)}{\leq} \frac{\Delta}{n C'(\Gamma)^2}.
     \label{finbhikarle}  
\end{align}
In inequality $(a)$, we used $(\ref{aaronerr2})$. In inequality $(b)$, we used $(\ref{aaronerr1})$ and $(\ref{bzuyv})$, noting that $\mathbb{E}[X'] = 0$. In inequality $(c)$, we used the inequality $\mathbb{E}[|X'|] \leq \sqrt{\mathbb{E}[X'^2]} = \sqrt{\text{Var}(X')} \leq \sqrt{V/n}$. In inequality $(d)$, we used the fact that $\Delta < 1/2$. Therefore, from $(\ref{finbhikarle})$, we have 
\begin{align}
    &\text{Var}\left( \frac{\sqrt{n} C'(\Gamma)}{\sqrt{\psi_\beta(c(\mathbf{X}))}} \left(\Gamma - c(\mathbf{X})  \right)\right) \notag \\
    &\leq \text{Var}\left( \frac{\sqrt{n} C'(\Gamma)}{\sqrt{V(\Gamma)}} \left(\Gamma - c(\mathbf{X}) \right)\right) + \Delta \notag  \\
    &\leq \frac{C'(\Gamma)^2 V}{V(\Gamma)}  + \Delta. \label{Piprimevar}
\end{align}

Define the random variable $\Pi'$
as 
\begin{align*}
    \Pi' \coloneqq  \frac{\sqrt{n} C'(\Gamma)}{\sqrt{\psi_\beta(c(\mathbf{X}))}} \left(\Gamma - c(\mathbf{X})  \right)
\end{align*}
so that, from $(\ref{Piprimemean})$ and $(\ref{Piprimevar})$, $\mathbb{E}[\Pi'] \geq -\Delta$ and $\text{Var}(\Pi') \leq \frac{C'(\Gamma)^2 V}{V(\Gamma)}  + \Delta$. Then, from $(\ref{pq})$, for sufficiently large $n$,
\begin{align}
    &(\overline{P} \circ W) \left( \log \frac{W(\mathbf{Y}|\mathbf{X})}{Q^*(\mathbf{Y})} > nC(\Gamma) + \sqrt{n} r \right) \notag \\
    &\leq 1 + 2 \frac{\kappa}{\sqrt{n}}  - \inf_{\Pi'} \mathbb{E}\left [ \Phi\left(\Pi' + \frac{ r}{\sqrt{\varphi(r)}} \right)\right ] \notag \\
    &\stackrel{(a)}{=} 1 + 2 \frac{\kappa}{\sqrt{n}} - \inf_{\Pi} \mathbb{E}\left [ \Phi(\Pi) \right ] \notag \\
    &\stackrel{(b)}{=} 1 + 2 \frac{\kappa}{\sqrt{n}} - \mathcal{K}\left( -\Delta +  \frac{ r}{\sqrt{\varphi(r)}}, \frac{C'(\Gamma)^2 V}{V(\Gamma)}  + \Delta \right). \label{iskosocrmain} 
\end{align}
The infimum in equality $(a)$ above is over all random variables $\Pi$ such that $\mathbb{E}[\Pi] \geq -\Delta +  \frac{ r}{\sqrt{\varphi(r)}}$ and $\text{Var}(\Pi) \leq \frac{C'(\Gamma)^2 V}{V(\Gamma)}  + \Delta.$ Equality $(b)$ follows by the definition and properties of the $\mathcal{K}$ function given in Definition \ref{Kfuncdef} and Lemma \ref{Kfuncproperties}.

Using $(\ref{iskosocrmain})$ to upper bound $(\ref{bx})$ and using that upper bound in the result of Lemma \ref{mostgenconv} (with $\rho = \exp\left(n C(\Gamma) + \sqrt{n} r \right)$), we obtain 
\begin{align}
    &\log M^*(n, \epsilon, \Gamma, V) \leq n C(\Gamma) + \sqrt{n}r - \log \left [ \left(  - \epsilon   -  2 \frac{\kappa}{\sqrt{n}} + \mathcal{K}\left( -\Delta +  \frac{ r}{\sqrt{\varphi(r)}}, \frac{C'(\Gamma)^2 V}{V(\Gamma)}  + \Delta \right)  \right)^+\right] \notag \\
    &\frac{\log M^*(n, \epsilon, \Gamma, V) - nC(\Gamma)}{\sqrt{n}} \leq r - \frac{1}{\sqrt{n}} \log \left [ \left(  - \epsilon   -  2 \frac{\kappa}{\sqrt{n}} + \mathcal{K}\left( -\Delta +  \frac{ r}{\sqrt{\varphi(r)}}, \frac{C'(\Gamma)^2 V}{V(\Gamma)}  + \Delta \right)  \right)^+\right]. \label{rchoosekar}
\end{align}
For any given average error probability $\epsilon \in (0, 1)$, we choose $r$ in $(\ref{rchoosekar})$ as 
\begin{align}
    r &= \begin{cases}
       r_1^* + \eta & \text{ if } \epsilon \in \left(0, \epsilon^* \right ] \\
       r_2^* + \eta & \text{ if } \epsilon \in \left(\epsilon^* , 1 \right ),
    \end{cases}
    \label{rfuncepsilon}
\end{align}
where  
\begin{align}
    r^*_1 &\coloneqq \max \left \{r' : \mathcal{K}\left(-\Delta + \frac{r'}{\sqrt{V(\Gamma) - \tau_\beta }}, \frac{C'(\Gamma)^2 V}{V(\Gamma)}  + \Delta \right) \leq \epsilon \right \}, \label{r1star}\\ 
    r^*_2 &\coloneqq \max \left \{r' : \mathcal{K}\left(-\Delta + \frac{r'}{\sqrt{V(\Gamma) + \tau_\beta }}, \frac{C'(\Gamma)^2 V}{V(\Gamma)}  + \Delta \right) \leq \epsilon \right \}, \label{r2star}\\
    \epsilon^* &\coloneqq \mathcal{K}\left(-\Delta - \frac{\eta}{\sqrt{V(\Gamma) - \tau_\beta}}, \frac{C'(\Gamma)^2 V}{V(\Gamma)}  + \Delta \right). \notag 
\end{align}
Note that in $(\ref{rfuncepsilon})$, $r \leq 0$ for $\epsilon \leq \epsilon^*$ and $r > 0$ for $\epsilon > \epsilon^*$.

Therefore, for $\epsilon \leq \epsilon^*$, we have 
\begin{align}
    \frac{\log M^*(n, \epsilon, \Gamma, V) - nC(\Gamma)}{\sqrt{n}} \leq r_1^* + \eta - \frac{1}{\sqrt{n}} \log \left [ \left(  - \epsilon   -  \frac{2\kappa}{\sqrt{n}} + \mathcal{K}\left( -\Delta +  \frac{ r_1^* + \eta}{\sqrt{V(\Gamma) - \tau_\beta}}, \frac{C'(\Gamma)^2 V}{V(\Gamma)}  + \Delta \right)  \right)^+\right].  \label{rchoosekarvv}
\end{align}
Since $\mathcal{K}(\cdot\,, \cdot)$ is strictly increasing in the first argument, we have 
\begin{align*}
    \mathcal{K}\left(-\Delta + \frac{r_1^* + \eta}{\sqrt{V(\Gamma) - \tau_\beta}}, \frac{C'(\Gamma)^2 V}{V(\Gamma)}  + \Delta \right) > \epsilon
\end{align*}
from the definition of $r_1^*$ in $(\ref{r1star})$ and the fact that $\eta > 0$. Hence, taking the limit supremum as $n \to \infty$ in $(\ref{rchoosekarvv})$, we obtain 
\begin{align}
    \limsup_{n \to \infty}\,\frac{\log M^*(n, \epsilon, \Gamma, V) - nC(\Gamma)}{\sqrt{n}} \leq r_1^* + \eta \label{pq2}
\end{align}
for $\epsilon \leq \epsilon^*$. For $\epsilon > \epsilon^*$, a similar derivation gives us 
\begin{align}
    \limsup_{n \to \infty}\,\frac{\log M^*(n, \epsilon, \Gamma, V) - nC(\Gamma)}{\sqrt{n}} \leq r_2^* + \eta. \label{pq3}
\end{align}
We now let $\Delta, \eta, \beta$ and $\tau_\beta$ go to zero in both $(\ref{pq2})$ and $(\ref{pq3})$. Then using the fact from Lemma \ref{Kfuncproperties} that $\mathcal{K}(\cdot, \cdot)$ is continuous, we obtain 
\begin{align*}
    \limsup_{n \to \infty}\,\frac{\log M^*(n, \epsilon, \Gamma, V) - nC(\Gamma)}{\sqrt{n}} \leq \max \left \{r : \mathcal{K}\left(\frac{r}{\sqrt{V(\Gamma) }}, \frac{C'(\Gamma)^2 V}{V(\Gamma)}   \right) \leq \epsilon \right \}
\end{align*}
for all $\epsilon \in (0, 1)$.

The converse result can also be stated in terms of a lower bound on the minimum average probability of error of $(n, R, \Gamma, V)$ codes for a rate $R = C(\Gamma) + \frac{r}{\sqrt{n}}$. Starting again from Lemma \ref{mostgenconv}, we have that for $(n, R, \Gamma, V)$ codes with minimum average error probability $\epsilon$, 
\begin{align}
    \log \lceil \exp(nR) \rceil &\leq \log \rho - \log \left [ \left( 1 - \epsilon - \sup_{\overline{P} \in \mathcal{P}_{\Gamma,V}(\mathbb{R}^n)}\, \inf_{q \in \mathcal{P}(\mathbb{R}^n)} (\overline{P} \circ W) \left( \frac{W(\mathbf{Y}|\mathbf{X})}{q(\mathbf{Y})} > \rho \right)\right)^+\right].\label{qandp22}
\end{align}
Assume first that $r \leq 0$ and let $\rho = \exp\left(n C(\Gamma) + r' \sqrt{n} \right)$ for some arbitrary $r' < r$. It directly follows from  
$(\ref{iskosocrmain})$ and the definition of $\varphi(\cdot)$ in $(\ref{varphidefinitionr})$ that
\begin{align}
    &\sup_{\overline{P} \in \mathcal{P}_{\Gamma,V}(\mathbb{R}^n)}\, \inf_{q \in \mathcal{P}(\mathbb{R}^n)} (\overline{P} \circ W) \left( \frac{W(\mathbf{Y}|\mathbf{X})}{q(\mathbf{Y})} > \rho \right) \notag \\
    &\leq  1 + 2 \frac{\kappa}{\sqrt{n}} - \mathcal{K}\left( -\Delta +  \frac{ r}{\sqrt{V(\Gamma) - \tau_\beta}}, \frac{C'(\Gamma)^2 V}{V(\Gamma)}  + \Delta \right) \label{nogames6}
\end{align}
From $(\ref{qandp22})$ and $(\ref{nogames6})$, we have 
\begin{align}
    \log \lceil \exp(nR) \rceil &\leq \log \rho - \log \left [ \left( - \epsilon -  2 \frac{\kappa}{\sqrt{n}} + \mathcal{K}\left(-\Delta +  \frac{r'}{\sqrt{V(\Gamma) - \tau_\beta}}, \frac{C'(\Gamma)^2 V}{V(\Gamma)}  + \Delta  \right)\right)^+\right] \label{z1s}
\end{align}
which evaluates to
\begin{align*}
     - \epsilon -  2 \frac{\kappa}{\sqrt{n}} + \mathcal{K}\left(-\Delta +  \frac{r'}{\sqrt{V(\Gamma) - \tau_\beta}}, \frac{C'(\Gamma)^2 V}{V(\Gamma)}  + \Delta  \right) \leq e^{-(r-r')\sqrt{n}}.
\end{align*}
Taking the limit as $n \to \infty$ and letting $r' \to r$, $\Delta \to 0$, $\beta \to 0$ and $\tau_\beta \to 0$, we obtain 
\begin{align}
    \epsilon \geq \mathcal{K}\left( \frac{r}{\sqrt{V(\Gamma) }}, \frac{C'(\Gamma)^2 V}{V(\Gamma)}   \right) \label{z1s2}
\end{align}
for any $r \leq 0$. For $r > 0$, let $\rho = \exp\left(n C(\Gamma) + r' \sqrt{n} \right)$ for some arbitrary $0 < r' < r$. Then from  
$(\ref{iskosocrmain})$ and the definition of $\varphi(\cdot)$ in $(\ref{varphidefinitionr})$, we have that  
\begin{align}
    &\sup_{\overline{P} \in \mathcal{P}_{\Gamma,V}(\mathbb{R}^n)}\, \inf_{q \in \mathcal{P}(\mathbb{R}^n)} (\overline{P} \circ W) \left( \frac{W(\mathbf{Y}|\mathbf{X})}{q(\mathbf{Y})} > \rho \right) \notag \\
    &\leq 1 +  2 \frac{\kappa}{\sqrt{n}} - \mathcal{K}\left(-\Delta +  \frac{r'}{\sqrt{V(\Gamma) + \tau_\beta}}, \frac{C'(\Gamma)^2 V}{V(\Gamma)}  + \Delta  \right). \label{nogddames6}
\end{align} 
Then a similar derivation to that used from $(\ref{z1s})$ to $(\ref{z1s2})$ gives us 
\begin{align}
    \epsilon \geq \mathcal{K}\left( \frac{r}{\sqrt{V(\Gamma) }}, \frac{C'(\Gamma)^2 V}{V(\Gamma)}   \right) \notag 
\end{align}
for all $r > 0$.

\end{IEEEproof}

\section{Future Directions}

In addition to the variance constraint $\operatorname{Var}(c(\mathbf{X})) \leq V/n$, higher moment constraints could be imposed on the block-average cost $c(\mathbf{X})$. Defining the normalized cost deviation $\tilde{c}(\mathbf{X}) \coloneqq \sqrt{n}(c(\mathbf{X}) - \Gamma)$, one could also study coding performance under the constraint specified by $\mathbb{E}[f(\tilde{c}(\mathbf{X}))] \leq \Delta$ for a broad class of functions $f : \mathbb{R} \to [0, \infty)$. A related direction is to study the scaling $\operatorname{Var}(c(\mathbf{X})) \leq V/n^s$ for a range of values of $s$ to model more complex random processes as channel inputs. This would also allow us to interpolate between weak concentration $(s < 1)$ and strong concentration $(s > 1)$ and to identify phase transitions in first/second-order asymptotics as $s$ varies. Other channel models could be studied under the m.v. cost constraint such as parallel Gaussian channels \cite{8012458}, \cite[Th. 78]{Polyanskiy2010}, multi-access channels \cite{7300429}, fading channels \cite{6736577}, \cite{6620483}, with and without feedback. The performance under the m.v. cost constraint could also be studied in the large deviations regime (error exponent analysis) and moderate deviations regime \cite{6814957}, \cite{10595462}.

\appendices

\section{Proof of Lemma \ref{sphericalsymm} \label{sphericalsymmproof}}

Let $Y \sim W(\cdot | x) = \mathcal{N}(x, N)$. We have  

\begin{align*}
    W(y|x) &= \frac{1}{\sqrt{2\pi N}} \exp\left(-\frac{(y-x)^2}{2N} \right)\\
    Q^*(y) &= \frac{1}{\sqrt{2\pi(\Gamma + N)}} \exp\left( -\frac{y^2}{2(\Gamma + N)} \right)\\
     \frac{W(Y|x)}{Q^*(Y)} &= \frac{\sqrt{\Gamma + N}}{\sqrt{ N}} \exp\left(-\frac{(Y-x)^2}{2N} + \frac{Y^2}{2(\Gamma + N)} \right)\\
    \log \frac{W(Y|x)}{Q^*(Y)} &= \frac{1}{2} \log \left(1 + \frac{\Gamma}{N} \right) -\frac{(Y-x)^2}{2N} + \frac{Y^2}{2(\Gamma + N)}\\
    \mathbb{E}\left [ \log \frac{W(Y|x)}{Q^*(Y)} \right] &= \frac{1}{2} \log \left(1 + \frac{\Gamma}{N} \right) -\frac{\Gamma - x^2}{2(\Gamma + N)}.
\end{align*}
Hence, 
\begin{align}
    T_i &= \frac{Y_i^2}{2(\Gamma + N)} -\frac{(Y_i-x_i)^2}{2N} + \frac{\Gamma - x_i^2}{2(\Gamma + N)}. 
\end{align}
Using the relation $Y_i = x_i + Z_i$, we can write  
\begin{align*}
   \sum_{i=1}^n T_i &= \sum_{i=1}^n \frac{(x_i + Z_i)^2}{2(\Gamma + N)} - \sum_{i=1}^n \frac{Z_i^2}{2N} + \frac{n\Gamma - n c(\mathbf{X})}{2(\Gamma + N)}\\
   &= \frac{n c(\mathbf{X})}{2(\Gamma + N)} + \sum_{i=1}^n \frac{Z_i^2}{2(\Gamma + N)} + \sum_{i=1}^n\frac{x_i Z_i}{\Gamma + N}  - \sum_{i=1}^n \frac{Z_i^2}{2N} + \frac{n\Gamma - n c(\mathbf{X})}{2(\Gamma + N)}\\
   &= \sum_{i=1}^n \left [ \frac{Z_i^2}{2(\Gamma + N)} + \frac{x_i Z_i}{\Gamma + N}  -  \frac{Z_i^2}{2N} \right] + \frac{n\Gamma }{2(\Gamma + N)}. 
\end{align*}
By completing the square and writing $Z_i = \sqrt{N} S_i$, where $\{S_i \}$ is i.i.d. $\mathcal{N}(0, 1)$, we can write  
\begin{align*}
    \sum_{i=1}^n T_i &= -\frac{\Gamma}{2(\Gamma + N)} \sum_{i=1}^n\left(S_i - x_i\frac{\sqrt{N}}{\Gamma} \right)^2 + \frac{N n c(\mathbf{X})}{2\Gamma(\Gamma + N)}  +  \frac{n\Gamma }{2(\Gamma + N)}.
\end{align*}
Since 
$$\sum_{i=1}^n\left(S_i - x_i\frac{\sqrt{N}}{\Gamma} \right)^2$$
has a noncentral chi-squared distribution with $n$ degrees of freedom and noncentrality parameter $\lambda$ given by 
\begin{align*}
    \lambda = \frac{N n c(\mathbf{X})}{\Gamma^2}, 
\end{align*}
the assertion of the lemma follows.

\section{Proof of Property 4 in Lemma \ref{Kfuncproperties} \label{prop4proof}}

We prove by contradiction. Suppose that there exists (property 1 of Lemma \ref{Kfuncproperties}) a minimizing probability distribution $P$ in $(\ref{min5})$ such that $\mathbb{E}_P[\Pi] = r$ and $\text{Var}_P(\Pi) = V' < V$. Let 
\begin{align*}
    P(\pi) = \begin{cases}
    p_1 & \pi = \pi_1\\
    p_2 & \pi = \pi_2\\
    p_3 & \pi = \pi_3.
    \end{cases}
\end{align*}
Without loss of generality, we assume $p_3 \neq 0$, $\pi_1 \leq \pi_2 \leq \pi_3$ and $\pi_3 > \pi_1$. If the minimizing distribution $P$ is a two-point mass distribution, then we assume that $p_2 = 0$ and $\pi_2$ is an arbitrary number between $\pi_1$ and $\pi_3$. Since $\mathbb{E}_P[\Phi(\Pi)] < \Phi(r)$ and $\Phi(\cdot)$ is convex over the interval $(-\infty, 0]$, we must have $\pi_3 > \max(r, 0)$. Fix any $a >0$ such that $V' + p_3 a^2 \leq V$ and $\pi_3 - a \geq 0$. Consider a modified distribution
\begin{align*}
    Q(\pi) = \begin{cases}
    p_1 & \pi = \pi_1\\
    p_2 & \pi = \pi_2\\
    \frac{p_3}{2} & \pi = \pi_3 - a\\
    \frac{p_3}{2} & \pi = \pi_3 + a.
    \end{cases}
\end{align*}
We have $\mathbb{E}_Q[\Pi] = r$, $\text{Var}_Q(\Pi) = V' + p_3 a^2  > V'$ and  
\begin{align*}
    &\mathbb{E}_{P}[\Phi(\Pi)] - \mathbb{E}_{Q}[\Phi(\Pi)]  \\
    &= p_3 \left [ \Phi(\pi_3) - \frac{1}{2} \Phi(\pi_3 - a) - \frac{1}{2} \Phi(\pi_3 + a) \right] > 0
\end{align*}
because $\Phi(\cdot)$ is concave over the interval $[\pi_3 - a, \pi_3 +a]$. 

Next, we define a connected, compact subset $\mathcal{S}^*$ of $\mathbb{R}^3$ as the codomain of the mapping $\zeta : I^* \to \mathcal{S}^*$, where $I^* = [\pi_1, \pi_3 + a]$ and 
\begin{align*}
    \zeta(\pi) = \left( \Phi(\pi), \pi, (\pi - r)^2 \right). 
\end{align*}
Then we note that 
\begin{align*}
     \mathbb{E}_{Q}\left [ \zeta(\Pi) \right ] = \left(\mathbb{E}_{Q} \left [  \Phi(\Pi)\right ] , \mathbb{E}_{Q}\left [ \Pi\right] , \mathbb{E}_{Q}\left [ (\Pi - r)^2\right] \right)
\end{align*}
is in the convex closure of $\mathcal{S}^*$. By Carathéodory–Fenchel's theorem \cite[Appendix A]{el_gamal_kim_2011}, $\mathbb{E}_{Q}\left [ \zeta(\Pi) \right ]$ can be written as the convex combination of at most $3$ points in $\mathcal{S}^*$. This implies that there exists a three-point probability distribution $\tilde{Q}$ such that 
\begin{itemize}
    \item $\mathbb{E}_{Q}[\Phi(\Pi)] = \mathbb{E}_{\tilde{Q}}[\Phi(\Pi)] < \mathbb{E}_{P}[\Phi(\Pi)]$, 
    \\
    \item $\mathbb{E}_{Q}[\Pi] = \mathbb{E}_{\tilde{Q}}[\Pi] = \mathbb{E}_{P}[\Pi] = r$, and \\
    
    \item $\text{Var}_{P} (\Pi) < \text{Var}_{Q}(\Pi) = \text{Var}_{\tilde{Q}}(\Pi) \leq V$. 
\end{itemize}
We thus have a contradiction.

\section{Proof of Lemma \ref{QccN} \label{QccNproof}}

Define $\tilde{\mathbf{Y}} = \tilde{\mathbf{X}} + \tilde{\mathbf{Z}}$, where $\tilde{\mathbf{X}}$ and $\tilde{\mathbf{Z}}$ are independent, $\tilde{\mathbf{X}}$ is uniformly distributed on an $(n-1)$ sphere of radius $\frac{R}{\sqrt{N}}$ and $\tilde{\mathbf{Z}} \sim \mathcal{N}(\mathbf{0}, I_n)$. Let $Q^{cc}_1$ denote the PDF of $\tilde{\mathbf{Y}}$. From \cite[Proposition 1]{8598797}, we have 
\begin{align*}
        Q^{cc}_1(\tilde{\mathbf{y}}) = \frac{\Gamma\left( \frac{n}{2} \right)}{2 \pi^{n/2}}  \cdot  \exp\left( -\frac{R^2 + N||\tilde{\mathbf{y}}||^2}{2N} \right)  \left( \frac{\sqrt{N}}{R||\tilde{\mathbf{y}}||}\right)^{\frac{n}{2}-1}  I_{\frac{n}{2}-1}\left( \frac{R||\tilde{\mathbf{y}}||}{\sqrt{N}}\right).
\end{align*}
Since $\mathbf{Y} \stackrel{d}{=} \sqrt{N} \tilde{\mathbf{Y}}$, we can apply the change-of-variables formula
\begin{align*}
    Q^{cc}(\mathbf{y}) &= Q_1^{cc}\left (\frac{1}{\sqrt{N}} \mathbf{y} \right) \cdot \text{det}\left( \frac{1}{\sqrt{N}} I_n \right)
\end{align*}
to obtain the result.

\section{Proof of Lemma \ref{Qccratiolemma} \label{Qccratiolemmaproof}}

 From Lemma \ref{QccN}, we have 
\begin{align*}
    Q^{cc}(\mathbf{y}) &= \frac{\Gamma\left( \frac{n}{2} \right)}{2 (\pi N)^{n/2}}  \cdot  \exp\left( -\frac{n\Gamma + ||\mathbf{y}||^2}{2N} \right)  \left( \frac{N}{\sqrt{n\Gamma}||\mathbf{y}||}\right)^{\frac{n}{2}-1}  I_{\frac{n}{2}-1}\left(\frac{\sqrt{n\Gamma}||\mathbf{y}||}{N}\right) 
\end{align*}
and 
\begin{align*}
    Q^{cc}_0(\mathbf{y}) &= \frac{\Gamma\left( \frac{n}{2} \right)}{2 (\pi N)^{n/2}}  \cdot  \exp\left( -\frac{n\Gamma' + ||\mathbf{y}||^2}{2N} \right)  \left( \frac{N}{\sqrt{n\Gamma'}||\mathbf{y}||}\right)^{\frac{n}{2}-1}  I_{\frac{n}{2}-1}\left(\frac{\sqrt{n\Gamma'}||\mathbf{y}||}{N}\right).
\end{align*}
Hence, 
\begin{align*}
    \frac{Q^{cc}_0(\mathbf{y})}{Q^{cc}(\mathbf{y})} &= \frac{\exp\left( -\frac{n\Gamma' }{2N} \right)  \left( \frac{N}{\sqrt{n\Gamma'}||\mathbf{y}||}\right)^{\frac{n}{2}-1}  I_{\frac{n}{2}-1}\left(\frac{\sqrt{n\Gamma'}||\mathbf{y}||}{N}\right)}{\exp\left( -\frac{n\Gamma }{2N} \right)  \left( \frac{N}{\sqrt{n\Gamma}||\mathbf{y}||}\right)^{\frac{n}{2}-1}  I_{\frac{n}{2}-1}\left(\frac{\sqrt{n\Gamma}||\mathbf{y}||}{N}\right)}\\
    &= \exp\left( \frac{n}{2N} (\Gamma - \Gamma') \right) \left( \frac{\sqrt{n\Gamma}||\mathbf{y}||}{\sqrt{n\Gamma'}||\mathbf{y}||}\right)^{\frac{n}{2}-1} \frac{    I_{\frac{n}{2}-1}\left(\frac{\sqrt{n\Gamma'}||\mathbf{y}||}{N}\right)}{ I_{\frac{n}{2}-1}\left(\frac{\sqrt{n\Gamma}||\mathbf{y}||}{N}\right)}\\
    &= \exp\left( \frac{n}{2N} (\Gamma - \Gamma') \right) \left( \frac{\Gamma}{\Gamma'}\right)^{\frac{n}{4}-\frac{1}{2}} \frac{ I_{\frac{n}{2}-1}\left(\frac{\sqrt{n\Gamma'}||\mathbf{y}||}{N}\right)}{ I_{\frac{n}{2}-1}\left(\frac{\sqrt{n\Gamma}||\mathbf{y}||}{N}\right)}.
\end{align*}
Hence, 
\begin{align*}
    &\log \frac{Q^{cc}_0(\mathbf{y})}{Q^{cc}(\mathbf{y})}\\
    &=   \frac{n}{2N} (\Gamma - \Gamma') + \left( \frac{n}{4} - \frac{1}{2} \right) \log \left(\frac{\Gamma}{\Gamma'} \right) + \log  I_{\frac{n}{2}-1}\left(\frac{\sqrt{n\Gamma'}||\mathbf{y}||}{N}\right) - \log I_{\frac{n}{2}-1}\left(\frac{\sqrt{n\Gamma}||\mathbf{y}||}{N}\right)
\end{align*}

To approximate the Bessel function, we first rewrite it as  
\begin{align*}
    I_{\frac{n}{2}-1}\left(\frac{\sqrt{n\Gamma}||\mathbf{y}||}{N}\right) &= I_{\nu}\left(\nu z  \right),
\end{align*}
where $\nu = \frac{n}{2} - 1$ and $z = \frac{2  \sqrt{n\Gamma} ||\mathbf{y}||}{N(n-2)}$. Since $\mathbf{y} \in \mathcal{P}_n^*$ by assumption, we have that $z$ lies in a compact interval $[a, b] \subset (0, \infty)$ for sufficiently large $n$, where $0 < a < b < \infty$. Hence, we can use a uniform asymptotic expansion of the modified Bessel function (see \cite[10.41.3]{NISTHandbook} whose interpretation is given in \cite[2.1(iv)]{NISTHandbook}): as $\nu \to \infty$ and for $0 < z < \infty$, we have 
\begin{align}
    I_{\nu}(\nu z) = \frac{e^{\nu \eta}}{(2 \pi \nu)^{1/2} (1 + z^2)^{\frac{1}{4}}} \left(1+ O \left(\frac{1}{\nu} \right) \right), \label{besselapprox22}
\end{align}
where
$$\eta = \sqrt{1 + z^2} + \log \left( \frac{z}{1 + (1 + z^2)^{1/2}}\right)$$
and the $O(1/\nu)$ term in $(\ref{besselapprox22})$ can be uniformly bounded over $\mathcal{P}_n^*$. Using the asymptotic expansion in $(\ref{besselapprox22})$, we have 
\begin{align}
    \log I_{\frac{n}{2}-1}\left(\frac{\sqrt{n\Gamma}||\mathbf{y}||}{N}\right) &= \frac{n}{2} \sqrt{1 + \frac{4 ||\mathbf{y}||^2 n\Gamma}{N^2(n-2)^2}} + \frac{n}{2} \log\left( \frac{2 ||\mathbf{y}|| \sqrt{n\Gamma}}{N(n-2)} \right) - \frac{n}{2} \log \left(1 + \sqrt{1 + \frac{4 ||\mathbf{y}||^2 n\Gamma}{N^2(n-2)^2}}  \right)   + O (\log n), \label{z322}
\end{align}
where it can be verified that the $O(\log n)$ term can be made to be uniformly bounded over $\mathcal{P}_n^*$. Using $(\ref{z322})$, we have 
\begin{align*}
    &\log  I_{\frac{n}{2}-1}\left(\frac{\sqrt{n\Gamma'}||\mathbf{y}||}{N}\right) -  I_{\frac{n}{2}-1}\left(\frac{\sqrt{n\Gamma}||\mathbf{y}||}{N}\right)\\
    &= \frac{n}{2} \sqrt{1 + \frac{4 ||\mathbf{y}||^2 n\Gamma'}{N^2(n-2)^2}} + \frac{n}{2} \log\left( \frac{2 ||\mathbf{y}|| \sqrt{n\Gamma'}}{N(n-2)} \right) - \frac{n}{2} \log \left(1 + \sqrt{1 + \frac{4 ||\mathbf{y}||^2 n\Gamma'}{N^2(n-2)^2}}  \right) - \mbox{}\\
    & \quad \quad \quad \quad \frac{n}{2} \sqrt{1 + \frac{4 ||\mathbf{y}||^2 n\Gamma}{N^2(n-2)^2}} - \frac{n}{2} \log\left( \frac{2 ||\mathbf{y}|| \sqrt{n\Gamma}}{N(n-2)} \right) + \frac{n}{2} \log \left(1 + \sqrt{1 + \frac{4 ||\mathbf{y}||^2 n\Gamma}{N^2(n-2)^2}}  \right) + O(\log n)\\
    &= \frac{n}{2} \sqrt{1 + \frac{4 ||\mathbf{y}||^2 n\Gamma'}{N^2(n-2)^2}} + \frac{n}{4} \log \left(\frac{\Gamma'}{\Gamma} \right) - \frac{n}{2} \log \left(1 + \sqrt{1 + \frac{4 ||\mathbf{y}||^2 n\Gamma'}{N^2(n-2)^2}}  \right) - \mbox{}\\
    & \quad \quad \quad \quad \frac{n}{2} \sqrt{1 + \frac{4 ||\mathbf{y}||^2 n\Gamma}{N^2(n-2)^2}}  + \frac{n}{2} \log \left(1 + \sqrt{1 + \frac{4 ||\mathbf{y}||^2 n\Gamma}{N^2(n-2)^2}}  \right) + O(\log n).
\end{align*}

Hence, 
\begin{align*}
    &\log \frac{Q^{cc}_0(\mathbf{y})}{Q^{cc}(\mathbf{y})}\\
    &= \frac{n}{2N} (\Gamma - \Gamma') + \left( \frac{n}{4} - \frac{1}{2} \right) \log \left(\frac{\Gamma}{\Gamma'} \right) + \mbox{}\\
    &\frac{n}{2} \sqrt{1 + \frac{4 ||\mathbf{y}||^2 n\Gamma'}{N^2(n-2)^2}} + \frac{n}{4} \log \left(\frac{\Gamma'}{\Gamma} \right) - \frac{n}{2} \log \left(1 + \sqrt{1 + \frac{4 ||\mathbf{y}||^2 n\Gamma'}{N^2(n-2)^2}}  \right) - \mbox{}\\
    & \quad \quad \quad \quad \frac{n}{2} \sqrt{1 + \frac{4 ||\mathbf{y}||^2 n\Gamma}{N^2(n-2)^2}}  + \frac{n}{2} \log \left(1 + \sqrt{1 + \frac{4 ||\mathbf{y}||^2 n\Gamma}{N^2(n-2)^2}}  \right) + O(\log n)\\
    &= \frac{n}{2N} (\Gamma - \Gamma')    + \frac{1}{2}  \log \left(\frac{\Gamma'}{\Gamma} \right) + \frac{n}{2} \sqrt{1 + \frac{4 ||\mathbf{y}||^2 n\Gamma'}{N^2(n-2)^2}}  - \frac{n}{2} \log \left(1 + \sqrt{1 + \frac{4 ||\mathbf{y}||^2 n\Gamma'}{N^2(n-2)^2}}  \right) - \mbox{}\\
    & \quad \quad \quad \quad \frac{n}{2} \sqrt{1 + \frac{4 ||\mathbf{y}||^2 n\Gamma}{N^2(n-2)^2}}  + \frac{n}{2} \log \left(1 + \sqrt{1 + \frac{4 ||\mathbf{y}||^2 n\Gamma}{N^2(n-2)^2}}  \right) + O(\log n). 
\end{align*}
To simplify the notation, we write 
$\delta = \frac{||\mathbf{y}||^2}{n}  - (\Gamma + N)$. Recall that $\Gamma' = \Gamma + \epsilon$ and $|\delta| \leq \Delta$. Then 
\begin{align}
    &\log \frac{Q^{cc}_0(\mathbf{y})}{Q^{cc}(\mathbf{y})} \notag \\
    &= -\frac{n}{2N} \epsilon    + \frac{1}{2} \log \left(1 + \frac{\epsilon}{\Gamma}  \right) + \frac{1}{2} g_n(\delta, \epsilon)    - \frac{1}{2} h_n(\delta, \epsilon) - \frac{1}{2} g_n(\delta, 0)  + \frac{1}{2} h_n(\delta, 0) + O(\log n), \label{plugbackin}
\end{align} 
where we define 
\begin{align*}
  g_n(\delta, \epsilon) &\coloneqq \sqrt{n^2 + \frac{4 \Gamma(\Gamma + N)}{N^2} \frac{n^4}{(n-2)^2} + \frac{4 \epsilon(\Gamma + N)}{N^2} \frac{n^4}{(n-2)^2} + \frac{4 \Gamma}{N^2} \frac{n^4}{(n-2)^2} \delta + \frac{4 \epsilon}{N^2} \frac{n^4}{(n-2)^2} \delta}\\
   h_n(\delta, \epsilon) &\coloneqq n \log \left(1 + \sqrt{1 + \frac{4  n^2 \Gamma (\Gamma + N + \delta)}{ N^2(n-2)^2} + \frac{4  n^2 \epsilon (\Gamma + N + \delta)}{ N^2(n-2)^2}}  \right).   
\end{align*}
Using the Taylor series approximation
\begin{align*}
    \frac{n^2}{(n-2)^2} = 1 + \frac{4}{n} + O\left( \frac{1}{n^2} \right), 
\end{align*}
we have 
\begin{align*}
&\left(g_n(\delta, \epsilon)\right)^2 \\
    &= n^2 + \frac{4 \Gamma(\Gamma + N)}{N^2} \left(n^2 + 4n + O(1)\right) + \frac{4 (\Gamma + N)}{N^2} \left(n^2 + 4n + O(1)\right)\epsilon + \mbox{}\\
    & \quad \quad \quad \quad \quad \quad \quad \frac{4 \Gamma}{N^2} \left(n^2 + 4n + O(1)\right) \delta + \frac{4 }{N^2} \left(n^2 + 4n + O(1)\right) \epsilon \delta\\
    &= n^2\left(1 + \frac{4 \Gamma(\Gamma + N)}{N^2} + \frac{4 (\Gamma + N)}{N^2} \epsilon + \frac{4\Gamma}{N^2} \delta + \frac{4}{N^2}\epsilon \delta \right) + \mbox{}\\
    & \quad \quad \quad \quad \quad \quad \quad  n \left( \frac{16 \Gamma(\Gamma + N)}{N^2} + \frac{16 (\Gamma + N)}{N^2} \epsilon + \frac{16\Gamma}{N^2} \delta + \frac{16}{N^2}\epsilon \delta  \right) + O(1). 
\end{align*}
Let 
\begin{align*}
    A &= 1 + \frac{4 \Gamma(\Gamma + N)}{N^2}\\
    B &=  \frac{4 (\Gamma + N)}{N^2}\\
    C &= \frac{4 \Gamma}{N^2}\\
    D &= \frac{4}{N^2} \\
    E &= \frac{16 \Gamma(\Gamma + N)}{N^2} 
\end{align*}
so that 
\begin{align*}
&g_n(\delta, \epsilon) \\
    &= \sqrt{n^2\left(A + B \epsilon + C \delta + D \epsilon \delta \right) + n \left(E + 4B \epsilon + 4 C \delta + 4D \epsilon \delta \right) + O(1)}\\
    &= n \sqrt{A + B \epsilon + C \delta + D \epsilon \delta + \frac{E + 4B \epsilon + 4 C \delta + 4D \epsilon \delta }{n} + O(n^{-2})}
\end{align*}
Further set $K = A + B \epsilon + C \delta + D \epsilon \delta$ and $L = E + 4B \epsilon + 4 C \delta + 4D \epsilon \delta$. Then 
\begin{align*}
g_n(\delta, \epsilon) &= n \sqrt{K + \frac{L }{n} + O(n^{-2})}\\
    &= n \sqrt{K} \sqrt{1 + \frac{L }{K n} + O(n^{-2})}\\
    &= n \sqrt{K} \left(1 + \frac{L}{2K n} + O(n^{-2}) \right)\\
    &= n \sqrt{K}  + \frac{L}{2 \sqrt{K}} + O(n^{-1}).
\end{align*}
Now 
\begin{align*}
    \sqrt{K} &= \sqrt{A + B \epsilon + C \delta + D \epsilon \delta}\\
    &= \sqrt{A} \sqrt{1 + \frac{B}{A} \epsilon + \frac{C}{A} \delta + \frac{D}{A} \epsilon \delta }\\
    &= \sqrt{A} \left(1 + \frac{B}{2A} \epsilon + \frac{C}{2A} \delta  + O(\epsilon^2, \Delta^2, \epsilon \Delta)  \right)\\
    &= \sqrt{A} + \frac{B}{2 \sqrt{A}} \epsilon + \frac{C}{2 \sqrt{A}} \delta + O(\epsilon^2, \Delta^2, \epsilon \Delta). 
\end{align*}
Thus 
\begin{align*}
    n \sqrt{K} &= n\sqrt{A} + \frac{B}{2 \sqrt{A}} n\epsilon + \frac{C}{2 \sqrt{A}}n \delta + O(n\epsilon^2, n\Delta^2, n\epsilon \Delta).
\end{align*}
Combining all the terms, we have 
\begin{align*}
    g_n(\delta, \epsilon) &= n \sqrt{1 + \frac{4 \Gamma(\Gamma + N)}{N^2}} +  \frac{4(\Gamma + N)}{2N^2 \sqrt{1 + \frac{4 \Gamma(\Gamma + N)}{N^2}}} n \epsilon + \frac{4\Gamma}{2N^2 \sqrt{1 + \frac{4 \Gamma(\Gamma + N)}{N^2}}} n \delta + O(1) + O(n\epsilon^2, n\Delta^2, n\epsilon \Delta).
\end{align*}
Note that 
\begin{align*}
   \sqrt{A}= \sqrt{1 + \frac{4 \Gamma(\Gamma + N)}{N^2}} &= \frac{2\Gamma + N}{N}. 
\end{align*}
Hence, 
\begin{align}
g_n(\delta, \epsilon) &= n \frac{2\Gamma + N}{N} +  \frac{2(\Gamma + N)}{N (2\Gamma + N) } n \epsilon + \frac{2\Gamma}{N(2\Gamma + N) } n \delta + O(1) + O(n\epsilon^2, n\Delta^2, n\epsilon \Delta). \label{gdeltaepsilon}
\end{align}

Now we turn to 
\begin{align*}
    h_n(\delta, \epsilon) = n \log \left(1 + \sqrt{1 + \frac{4  n^2 \Gamma (\Gamma + N + \delta)}{ N^2(n-2)^2} + \frac{4  n^2 \epsilon (\Gamma + N + \delta)}{ N^2(n-2)^2}}  \right).
\end{align*}
We first simplify the expression inside the square root as follows: 

\begin{align*}
    &1 + \frac{4  n^2 \Gamma (\Gamma + N + \delta)}{ N^2(n-2)^2} + \frac{4  n^2 \epsilon (\Gamma + N + \delta)}{ N^2(n-2)^2}\\
    &= 1 + \frac{4   \Gamma (\Gamma + N + \delta)}{ N^2} \left(1 + \frac{4}{n} + O\left(\frac{1}{n^2} \right) \right) + \frac{4  \epsilon (\Gamma + N + \delta)}{ N^2}  \left(1 + \frac{4}{n} + O\left(\frac{1}{n^2} \right) \right) \\
    &= 1 + \frac{4   \Gamma (\Gamma + N + \delta)}{ N^2} + \frac{4  \epsilon (\Gamma + N + \delta)}{ N^2} + O\left( \frac{1}{n} \right)\\
    &= A + C \delta + B \epsilon + D \epsilon \delta + O(n^{-1})\\
    &= A\left(1 + \frac{C}{A} \delta + \frac{B}{A} \epsilon + \frac{D}{A} \epsilon \delta + O(n^{-1})  \right).
\end{align*}
Therefore, 
\begin{align*}
    &\sqrt{A}\sqrt{1 + \frac{C}{A} \delta + \frac{B}{A} \epsilon + \frac{D}{A} \epsilon \delta + O(n^{-1})}\\
    &= \sqrt{A} \left(1 + \frac{C}{2A} \delta + \frac{B}{2A} \epsilon  + O(n^{-1}, \Delta^2, \epsilon^2, \epsilon \Delta) \right)\\
    &= \sqrt{A} + \frac{C}{2\sqrt{A}} \delta + \frac{B}{2\sqrt{A}} \epsilon  + O(n^{-1}, \Delta^2, \epsilon^2, \epsilon \Delta).
\end{align*}
Therefore, 
\begin{align*}
    &\log \left(1 + \sqrt{A} + \frac{C}{2\sqrt{A}} \delta + \frac{B}{2\sqrt{A}} \epsilon  + O(n^{-1}, \Delta^2, \epsilon^2, \epsilon \Delta)  \right)\\
    &= \log \left(1 + \sqrt{A} \right) + \frac{1}{1 + \sqrt{A}}\left( \frac{C}{2\sqrt{A}} \delta + \frac{B}{2\sqrt{A}} \epsilon \right) + O(n^{-1}, \Delta^2, \epsilon^2, \epsilon \Delta)\\
    &= \log \left(2 \frac{\Gamma + N}{N} \right) + \frac{N}{2(\Gamma + N)}\frac{4\Gamma}{2N(2\Gamma + N) } \delta + \frac{N}{2(\Gamma + N)} \frac{4(\Gamma + N)}{2N(2 \Gamma + N)} \epsilon + O(n^{-1}, \Delta^2, \epsilon^2, \epsilon \Delta)\\
    &= \log \left(2 \frac{\Gamma + N}{N} \right) + \frac{\Gamma}{(2\Gamma + N)(\Gamma + N) } \delta +  \frac{1}{2 \Gamma + N} \epsilon + O(n^{-1}, \Delta^2, \epsilon^2, \epsilon \Delta).
\end{align*}
Finally, 
\begin{align}
    h_n(\delta, \epsilon) &= n\log \left(2 \frac{\Gamma + N}{N} \right) + \frac{\Gamma}{(2\Gamma + N)(\Gamma + N) } n \delta +  \frac{1}{2 \Gamma + N} n\epsilon + O(n \Delta^2, n\epsilon^2, n\epsilon \Delta) + O(1). \label{hdeltaepsilon}
\end{align}

Substituting $(\ref{gdeltaepsilon})$ and $(\ref{hdeltaepsilon})$ in $(\ref{plugbackin})$, we obtain  
\begin{align*}
    &\log \frac{Q^{cc}_0(\mathbf{y})}{Q^{cc}(\mathbf{y})}\\
    &= -\frac{n}{2N} \epsilon    + \frac{1}{2} \log \left(1 + \frac{\epsilon}{\Gamma}  \right) + \frac{1}{2} g_n(\delta, \epsilon)    - \frac{1}{2} h_n(\delta, \epsilon) - \frac{1}{2} g_n(\delta, 0)  + \frac{1}{2} h_n(\delta, 0) + O(\log n)\\
    &= -\frac{n}{2N} \epsilon    + \frac{1}{2} \log \left(1 + \frac{\epsilon}{\Gamma}  \right) + \frac{1}{2} \left( n \frac{2\Gamma + N}{N} +  \frac{2(\Gamma + N)}{N (2\Gamma + N) } n \epsilon + \frac{2\Gamma}{N(2\Gamma + N) } n \delta + O(1) + O(n\epsilon^2, n\Delta^2, n\epsilon \Delta)  \right)    - \mbox{} \\
    &  \quad \quad \quad \frac{1}{2} \left( n\log \left(2 \frac{\Gamma + N}{N} \right) + \frac{\Gamma}{(2\Gamma + N)(\Gamma + N) } n \delta +  \frac{1}{2 \Gamma + N} n\epsilon + O(n \Delta^2, n\epsilon^2, n\epsilon \Delta) + O(1) \right) + O(\log n)  - \mbox{}\\
    &\quad \quad \quad \quad \quad \quad \quad  \frac{1}{2} \left( n \frac{2\Gamma + N}{N}  + \frac{2\Gamma}{N(2\Gamma + N) } n \delta + O(1) + O( n\Delta^2) \right)  + \mbox{} \notag \\
    & \quad \quad \quad \quad \quad \quad \quad \quad \quad \frac{1}{2}\left(  n\log \left(2 \frac{\Gamma + N}{N} \right) + \frac{\Gamma}{(2\Gamma + N)(\Gamma + N) } n \delta + O(n \Delta^2) + O(1) \right)\\
    &= -\frac{n}{2N} \epsilon + \frac{1}{2} \frac{2(\Gamma + N)}{N (2\Gamma + N) } n \epsilon + O(1) + O(n\epsilon^2, n\Delta^2, n\epsilon \Delta) - \frac{1}{2} \frac{1}{2 \Gamma + N} n\epsilon + O(\log n)\\
    &= O(\log n) + O(n\epsilon^2, n\Delta^2, n\epsilon \Delta).
\end{align*}

\section{Proof of Lemma \ref{Logn_lemma} \label{Logn_lemmaproof}}

From Lemma \ref{QccN}, we have 
\begin{align*}
    Q^{cc}(\mathbf{y}) &= \frac{\Gamma\left( \frac{n}{2} \right)}{2 (\pi N)^{n/2}}  \cdot  \exp\left( -\frac{n\Gamma + ||\mathbf{y}||^2}{2N} \right)  \left( \frac{N}{\sqrt{n\Gamma}||\mathbf{y}||}\right)^{\frac{n}{2}-1}  I_{\frac{n}{2}-1}\left(\frac{\sqrt{n\Gamma}||\mathbf{y}||}{N}\right).
\end{align*}
From the standard formula for the multivariate Gaussian, we have 
\begin{align*}
    Q^*(\mathbf{y}) &= \frac{1}{(2\pi(\Gamma + N))^{n/2}} \exp\left(- \frac{1}{2(\Gamma + N) } ||\mathbf{y}||^2 \right).
\end{align*}
Then 
\begin{align}
&\log \frac{Q^{cc}(\mathbf{y})}{Q^*(\mathbf{y})} \notag \\
    &= \log \left( \Gamma\left( \frac{n}{2} \right) \right) -  \frac{n\Gamma + ||\mathbf{y}||^2}{2N} + \frac{n}{2} \log\left(  \frac{N}{\sqrt{n\Gamma}||\mathbf{y}||} \right) - \log\left(  \frac{N}{\sqrt{n\Gamma}||\mathbf{y}||} \right) + \log \left( I_{\frac{n}{2}-1}\left(\frac{\sqrt{n\Gamma}||\mathbf{y}||}{N}\right) \right) \notag \\
    & \quad \quad \quad \quad \quad - \log(2) - \frac{n}{2} \log(\pi N) + \frac{n}{2}\log (2\pi(\Gamma + N)) + \frac{1}{2(\Gamma + N)} ||\mathbf{y}||^2 \notag \\
    &= \log \left( \Gamma\left( \frac{n}{2} \right) \right) -  \frac{ ||\mathbf{y}||^2 \Gamma }{2N(\Gamma + N)} - \frac{n\Gamma}{2N} + \frac{n}{2} \log\left(  \frac{N}{\sqrt{n\Gamma}||\mathbf{y}||} \right) - \log\left(  \frac{N}{\sqrt{n\Gamma}||\mathbf{y}||} \right) + \log \left( I_{\frac{n}{2}-1}\left(\frac{\sqrt{n\Gamma}||\mathbf{y}||}{N}\right) \right) \notag \\
    &  \quad \quad \quad \quad \quad - \log(2)  + \frac{n}{2} \log \left( \frac{2\Gamma + 2N}{N}\right) \notag  \\
    &\stackrel{(a)}{=} \frac{n}{2} \log\left(\frac{n}{2} \right) - \frac{n}{2} - \frac{1}{2} \log\left(\frac{n}{2} \right) + O(1) -  \frac{ ||\mathbf{y}||^2 \Gamma }{2N(\Gamma + N)} - \frac{n\Gamma}{2N} + \frac{n}{2} \log\left(  \frac{N}{\sqrt{n\Gamma}||\mathbf{y}||} \right) - \log\left(  \frac{N}{\sqrt{n\Gamma}||\mathbf{y}||} \right) + \mbox{} \notag \\
    &  \quad \quad \quad \quad \quad   \log \left( I_{\frac{n}{2}-1}\left(\frac{\sqrt{n\Gamma}||\mathbf{y}||}{N}\right) \right) - \log(2) + \frac{n}{2} \log \left( \frac{2\Gamma + 2N}{N}\right) \notag \\
    &= \frac{n}{2} \log\left(n \right) - \frac{n}{2} - \frac{1}{2} \log\left(\frac{n}{2} \right)  -  \frac{ ||\mathbf{y}||^2 \Gamma }{2N(\Gamma + N)} - \frac{n\Gamma}{2N} + \frac{n}{2} \log\left(  \frac{N}{\sqrt{n\Gamma}||\mathbf{y}||} \right) - \log\left(  \frac{N}{\sqrt{n\Gamma}||\mathbf{y}||} \right) + \mbox{} \notag \\
    &  \quad \quad \quad \quad \quad   \log \left( I_{\frac{n}{2}-1}\left(\frac{\sqrt{n\Gamma}||\mathbf{y}||}{N}\right) \right)  + \frac{n}{2} \log \left( \frac{\Gamma + N}{N}\right) + O(1) \label{b20}
\end{align}
In equality $(a)$, we used an asymptotic expansion of the log gamma function (see, e.g., \cite[5.11.1]{NISTHandbook}). To approximate the Bessel function, we first rewrite it as  
\begin{align*}
    I_{\frac{n}{2}-1}\left(\frac{\sqrt{n\Gamma}||\mathbf{y}||}{N}\right) &= I_{\nu}\left(\nu z  \right),
\end{align*}
where $\nu = \frac{n}{2} - 1$ and $z = \frac{2  \sqrt{n\Gamma} ||\mathbf{y}||}{N(n-2)}$. Since $\mathbf{y} \in \mathcal{P}_n^*$ by assumption, we have that $z$ lies in a compact interval $[a, b] \subset (0, \infty)$ for sufficiently large $n$, where $0 < a < b < \infty$. Hence, we can use a uniform asymptotic expansion of the modified Bessel function (see \cite[10.41.3]{NISTHandbook} whose interpretation is given in \cite[2.1(iv)]{NISTHandbook}): as $\nu \to \infty$ and for $0 < z < \infty$, we have 
\begin{align}
    I_{\nu}(\nu z) = \frac{e^{\nu \eta}}{(2 \pi \nu)^{1/2} (1 + z^2)^{\frac{1}{4}}} \left(1+ O \left(\frac{1}{\nu} \right) \right), \label{besselapprox}
\end{align}
where
$$\eta = \sqrt{1 + z^2} + \log \left( \frac{z}{1 + (1 + z^2)^{1/2}}\right).$$
and the $O(1/\nu)$ term in $(\ref{besselapprox})$ can be uniformly bounded over $\mathcal{P}_n^*$. Using the asymptotic expansion in $(\ref{besselapprox})$, we have 
\begin{align}
    \log I_{\frac{n}{2}-1}\left(\frac{\sqrt{n\Gamma}||\mathbf{y}||}{N}\right) &= \frac{n}{2} \sqrt{1 + \frac{4 ||\mathbf{y}||^2 n\Gamma}{N^2(n-2)^2}} + \frac{n}{2} \log\left( \frac{2 ||\mathbf{y}|| \sqrt{n\Gamma}}{N(n-2)} \right) - \frac{n}{2} \log \left(1 + \sqrt{1 + \frac{4 ||\mathbf{y}||^2 n\Gamma}{N^2(n-2)^2}}  \right)   + O (\log n), \label{z3}
\end{align}
where it can be verified that the $O(\log n)$ term can be made to be uniformly bounded over $\mathcal{P}_n^*$. Substituting $(\ref{z3})$ in $(\ref{b20})$, we obtain  
\begin{align}
&\log \frac{Q^{cc}(\mathbf{y})}{Q^*(\mathbf{y})} \notag \\
    &= \frac{n}{2} \log\left(n \right) - \frac{n}{2}  -  \frac{ ||\mathbf{y}||^2 \Gamma }{2N(\Gamma + N)} - \frac{n\Gamma}{2N} + \frac{n}{2} \log\left(  \frac{N}{\sqrt{n\Gamma}||\mathbf{y}||} \right) - \log\left(  \frac{N}{\sqrt{n\Gamma}||\mathbf{y}||} \right) +  \frac{n}{2} \sqrt{1 + \frac{4 ||\mathbf{y}||^2 n\Gamma}{N^2(n-2)^2}} + \mbox{}  \notag \\
    &  \quad \quad \quad \quad \quad  \frac{n}{2} \log\left( \frac{2 ||\mathbf{y}|| \sqrt{n\Gamma}}{N(n-2)} \right) - \frac{n}{2} \log \left(1 + \sqrt{1 + \frac{4 ||\mathbf{y}||^2 n\Gamma}{N^2(n-2)^2}}  \right) + \frac{n}{2} \log \left(\frac{\Gamma + N}{N} \right)  + O (\log n) \notag  \\
    &=  - \frac{n}{2}  -  \frac{ ||\mathbf{y}||^2 \Gamma }{2N(\Gamma + N)} - \frac{n\Gamma}{2N}   +  \frac{n}{2} \sqrt{1 + \frac{4 ||\mathbf{y}||^2 n\Gamma}{N^2(n-2)^2}} - \frac{n}{2} \log \left(1 + \sqrt{1 + \frac{4 ||\mathbf{y}||^2 n\Gamma}{N^2(n-2)^2}}  \right) \mbox{}  \notag \\
    &  \quad \quad \quad \quad \quad    + \frac{n}{2} \log \left(\frac{2\Gamma + 2N}{N} \right) + \log\left( \sqrt{n\Gamma}||\mathbf{y}|| \right) + O (\log n) \notag \\
    &= \left [ \frac{1}{2} g_n(\delta, 0) - \frac{\Gamma}{N} n - \frac{n}{2} - \frac{\Gamma}{2N(\Gamma + N)} n \delta \right] + \left [ \frac{n}{2} \log \left( \frac{2\Gamma + 2 N}{N} \right) - \frac{1}{2} h_n(\delta, 0) \right] +  \log\left( \sqrt{n\Gamma}||\mathbf{y}|| \right) + O (\log n), \label{42v}
\end{align}
where in the last equality above, we have  
\begin{align}
    \delta &= \frac{||\mathbf{y}||^2}{n}  - (\Gamma + N) \label{del_Def}\\
    g_n(\delta, 0) &= \sqrt{n^2 + \frac{4 \Gamma(\Gamma + N)}{N^2} \frac{n^4}{(n-2)^2} + \frac{4\Gamma}{N^2} \frac{n^4}{(n-2)^2} \delta} \notag \\
    h_n(\delta, 0) &= n \log \left(1 + \sqrt{1 + \frac{4  n^2 \Gamma (\Gamma + N + \delta)}{ N^2(n-2)^2}}  \right) \notag 
\end{align}
to facilitate further analysis. Recall that $|\delta| \leq \Delta$. 

From $(\ref{gdeltaepsilon})$, we have 
\begin{align*}
    g_n(\delta, 0) &= n\frac{2 \Gamma + N}{N} + \frac{2\Gamma}{N(2\Gamma + N)} n \delta + O(n \Delta^2) + O\left( 1 \right).
\end{align*}
Thus, we can simplify the first term in square brackets in $(\ref{42v})$ as  
\begin{align}
    &\frac{1}{2} g_n(\delta) -  \frac{\Gamma}{N} n - \frac{n}{2} - \frac{\Gamma}{2N(\Gamma + N)} n \delta \notag \\
     &= \frac{\Gamma}{N(2\Gamma + N)} n \delta + O(n \Delta^2)   + O\left( 1 \right)  - \frac{\Gamma}{2N(\Gamma + N)}n\delta \notag \\
     &= \frac{\Gamma}{2(2\Gamma + N)(\Gamma + N)} n \delta + O(n \Delta^2)  + O\left( 1 \right). \label{firstterm_sq}
\end{align}
Similarly, from $(\ref{hdeltaepsilon})$, we have
\begin{align*}
    h_n(\delta, 0) &=  n \log \left( \frac{2 \Gamma + 2N}{N} \right) + \frac{ \Gamma}{ (2 \Gamma + N)(\Gamma + N)} n\delta   + O(1) + O\left( n \Delta^2 \right). 
\end{align*}
Thus, we can simplify the second term in square brackets in $(\ref{42v})$ as 
\begin{align}
     &\frac{n}{2}\log \left(\frac{2\Gamma + 2N}{N} \right) - \frac{n}{2} \log \left(1 + \sqrt{1 + \frac{4  n^2 \Gamma (\Gamma + N + \delta)}{ N^2(n-2)^2}}  \right) \notag \\
     &=  - \frac{ \Gamma}{ 2(2 \Gamma + N)(\Gamma + N)} n\delta  + O(1) + O\left( n \Delta^2\right). \label{secondterm_sq}
\end{align}
Adding $(\ref{firstterm_sq})$ and $(\ref{secondterm_sq})$ gives us $O(n \Delta^2) + O(1)$. Hence, going back to $(\ref{42v})$, we have 
\begin{align*}
    &\log \frac{Q^{cc}(\mathbf{y})}{Q^*(\mathbf{y})}\\
    &= \log\left( \sqrt{n\Gamma}||\mathbf{y}|| \right) + O(\log n) + O(n\Delta^2)\\
    &= O(\log n) + O(n\Delta^2),
\end{align*}
where the last equality follows from the assumption that $\mathbf{y} \in \mathcal{P}_n^*$.

\section*{Acknowledgment}

This research was supported by the US National Science Foundation under grant CCF-1956192.


\ifCLASSOPTIONcaptionsoff
  \newpage
\fi



\bibliographystyle{IEEEtran}


%








\end{document}